\documentclass[5p,times,twocolumn]{elsarticle}
\usepackage{lineno,hyperref}
\usepackage{subcaption}
\usepackage{tikz}
\usepackage{pgfplotstable}
\usepackage{pgfplots}
\usepackage{multirow}
\usepackage{float}
\usepackage{textcomp}
\usepackage{amsmath,amsfonts,amssymb,bbm}	
\usepackage{multicol}
\usepackage{graphicx}
\usepackage{gensymb}
\usepackage{color,soul}
\usepackage{cleveref}[2012/02/15]

\usepackage[section]{placeins}

\modulolinenumbers[5]

\crefformat{footnote}{#2\footnotemark[#1]#3}

\usepackage{multirow} 
\usepackage{booktabs} 
\usepackage{placeins} 
\usepackage{hyperref}
\usepackage{rotating} 

\begin{document}

\begin{frontmatter}

\title{ReconResNet: Regularised Residual Learning for MR Image Reconstruction of Undersampled Cartesian and Radial Data}

\author[1,2,3,4]{Soumick Chatterjee\corref{mycorrespondingauthor}}
\cortext[mycorrespondingauthor]{Corresponding author}
\ead{soumick.chatterjee@ovgu.de}

\author[3,4]{Mario Breitkopf}
\author[3,4,5]{Chompunuch Sarasaen}
\author[5]{Hadya Yassin}
\author[4,5]{Georg Rose}
\author[1,2,7]{Andreas N{\"u}rnberger}
\author[3,4,6,7,8]{Oliver~Speck}

\address[1]{Faculty of Computer Science, Otto von Guericke University Magdeburg, Germany}
\address[2]{Data and Knowledge Engineering Group, Otto von Guericke University Magdeburg, Germany}
\address[3]{Biomedical Magnetic Resonance, Otto von Guericke University Magdeburg, Germany}
\address[4]{Research Campus STIMULATE, Otto von Guericke University Magdeburg, Germany}
\address[5]{Institute for Medical Engineering, Otto von Guericke University Magdeburg, Germany}
\address[6]{German Center for Neurodegenerative Disease, Magdeburg, Germany}
\address[7]{Center for Behavioral Brain Sciences, Magdeburg, Germany}
\address[8]{Leibniz Institute for Neurobiology, Magdeburg, Germany}

\begin{abstract}
MRI is an inherently slow process, which leads to long scan time for high-resolution imaging. The speed of acquisition can be increased by ignoring parts of the data (undersampling). Consequently, this leads to the degradation of image quality, such as loss of resolution or introduction of image artefacts. This work aims to reconstruct highly undersampled Cartesian or radial MR acquisitions, with better resolution and with less to no artefact compared to conventional techniques like compressed sensing. In recent times, deep learning has emerged as a very important area of research and has shown immense potential in solving inverse problems, e.g. MR image reconstruction. In this paper, a deep learning based MR image reconstruction framework is proposed, which includes a modified regularised version of ResNet as the network backbone to remove artefacts from the undersampled image, followed by data consistency steps that fusions the network output with the data already available from undersampled k-space in order to further improve reconstruction quality. The performance of this framework for various undersampling patterns has also been tested, and it has been observed that the framework is robust to deal with various sampling patterns, even when mixed together while training, and results in very high quality reconstruction, in terms of high SSIM (highest being 0.990$\pm$0.006 for acceleration factor of 3.5), while being compared with the fully sampled reconstruction. It has been shown that the proposed framework can successfully reconstruct even for an acceleration factor of 20 for Cartesian (0.968$\pm$0.005) and 17 for radially (0.962$\pm$0.012) sampled data. Furthermore, it has been shown that the framework preserves brain pathology during reconstruction while being trained on healthy subjects.
\end{abstract}

\begin{keyword}
MRI\sep MR Image Reconstruction\sep Undersampled MRI\sep Undersampled MR Reconstruction\sep Radial sampling reconstruction\sep Deep Learning
\end{keyword}

\end{frontmatter}


\section{Introduction}
\label{sec:introduction}
Magnetic resonance imaging (MRI) can provide high spatial resolution for detecting minute pathological changes in tissues. However, due to consecutive data acquisition, MRI is an inherently slow process~\citep{mcrobbie2017mri}. Fast imaging can improve patient compliance, reduce motion artefacts, increase patient throughput etc. MRI data is not acquired in spatial but frequency domain, which is the 2D Fourier transform of the image, the so-called k-space. To get back to image space a 2D inverse Fourier transform has to be applied. The speed of acquisition can be increased by ignoring parts of the k-space (undersampling). Taking the inverse Fourier transform from not densely enough sampled k-space frequency data might cause the resultant image to lose resolution and might also have artefacts due to the violation of the Nyquist-Shannon sampling theorem~\citep{nyquist1928certain,shannon1949communication}. Many of the approaches that are currently available for the reconstruction of undersampled data are very slow, due to the fact they are very computationally heavy or iterative in nature. For example, compressed sensing techniques are successfully applied to accelerate MR image acquisition~\citep{he2020dynamic,ueda2021compressed}, however, it expands reconstruction time and may result in incoherent artefacts or unrealistic images ~\citep{donoho2006compressed, Lustig.2007, knoll2011second}. Many of the established methods such as parallel imaging (sensitivity encoding - SENSE technique~\citep{pruessmann1999sense} or Generalised Auto-calibrating Partially Parallel Acquisitions - GRAPPA~\citep{griswold2002generalized}), which are not necessarily slow, can also help reduce the scan-time. This research work tries to complement parallel imaging techniques.

This paper proposes a robust reconstruction technique using deep learning for highly undersampled Cartesian or radial sampled data, which can reconstruct the image with better resolution and with less to no artefact, compared to classical methods. Generally, for deep learning based MR reconstruction~\citep{ChangMinHyun.2018}, the same sampling pattern is always applied to training as well as on testing data. In this work, different networks were trained on individual sampling patterns, as well as combined training was performed.

\subsection{Background and Related Work}\label{sec:Related}
Deep learning has proven to be a powerful tool for a wide array of applications. This approach has been utilised widely for MR image reconstruction or to reduce motion artefacts~\citep{qin2018convolutional,lyu2020cine}. In order to accelerate MRI acquisition, \citet{wang2016accelerating} and \citet{hammernik2018learning} applied deep learning to compressed sensing  MRI. Moreover, \cite{kwon2017parallel} proposed the multi-layer perception algorithm to reconstruct MRI as well as reducing aliasing artefacts for parallel imaging.

In 2016, deep residual learning; ResNet~\citep{He.2016} was proposed, to improve the accuracy while facilitating the optimisation of the deep neural network architecture. This enabled a pathway for deeper networks, without facing the vanishing gradient problem. It was initially examined for image recognition, by reassigning the layers for learning residual functions corresponding to the layer inputs. Since then, ResNet has been used for many applications such as image classification~\citep{mou2017unsupervised,zhang2019attention}, image denoising~\citep{zhang2017beyond,jifara2019medical}, Image-to-Image translation~\citep{Zhu.30032017}, and image segmentation~\citep{pakhomov2019deep}. Moreover, residual learning-based methods have been demonstrated to be an effective model to cope with undersampled MRI reconstruction problems as well~\citep{lee2017deep,lee2018deep,mardani2017deep,yao2019dr2}. 

To mitigate the long scan time by acquiring only parts of data in k-space (frequency domain), different undersampling patterns have been proposed over the years. Cartesian sampling or spin warp imaging is the most conventional way of sampling k-space~\citep{edelstein1980spin}. Random or pseudo-random sampling in the Cartesian acquisition is one of the types of undersampling often used in projects addressing the reconstruction of sparse k-space data~\citep{Lustig.2007,haldar2010compressed,hammernik2018learning}. Uniform undersampling is another popular way of undersampling data in Cartesian acquisitions~\citep{griswold2002generalized,akccakaya2019scan}. However, Cartesian acquisitions might be corrupted by image artefacts caused by respiratory or patient motion if not handled explicitly by the acquisition sequence~\citep{maclaren2013prospective}. Nevertheless, there are many different ways to sample the k-space and non-Cartesian samplings such as Radial sampling might be able to cope with this problem of motion artefacts by a repetitive sampling of the high-energy k-space centre~\citep{meyer_cartesian_2017}.  Various techniques have been proposed to reconstruct undersampled radial k-space~\citep{block2007undersampled,uecker2010nonlinear}. One interesting question in radial sampling is the way to optimally choose the angle between the spokes of acquisition. One sophisticated solution is to increment each spoke by $111.25^\circ$, which is $180^\circ$ divided by the golden ratio ($\varphi=\frac{1+\sqrt{5}}{2}\approx1.618$), known as golden angle radial sampling~\citep{Feng.2014,Block.2014}. Owing to the influence of the undersampling pattern on the reconstructed images, several studies have investigated the optimal undersampling trajectory of MRI~\citep{seeger2010optimization,haldar2019oedipus,bahadir2019learning}.

\subsection{Contributions}\label{sec:Contrib}
This paper proposes an MR reconstruction framework for undersampled Cartesian and radial data. The paper further proposes a ResNet-based model with modified residual blocks as the network backbone of the framework. Apart from the network, the framework uses an existing data consistency step for Cartesian sampling and introduces one for radial sampling, to replace all the real acquired data-points from the undersampled k-space in the network's output, to generate the final output of the framework. The framework was evaluated on two benchmark brain MRI datasets of healthy subjects and two datasets with pathology. Furthermore, different levels and types of undersampling have been tested with the framework. Additional experiments were performed by training different sampling patterns together and also by reconstructing different orientations than used for training.


\section{Methodology}\label{sec:Methodology}
The proposed framework, codenammed NCC1701, contains two main components: the network backbone architecture - ReconResNet and the data consistency step. This section presents the initial assumptions of the authors selecting ReconResNet for the task of undersampled MR reconstruction, then explains the architecture of the proposed ReconResNet, elucidates the different data consistency steps for Cartesian and radial sampling, elaborates about the implementation, presents the data used in this research, discusses the undersampling patterns and finally describes the evaluation techniques.

\subsection{Model Ansatz}
According to the universal approximation theorem~\citep{K.Hornik.1991}, it should be possible to find a nonlinear function $\Psi(x)$ that maps the corrupted input $x$ to the artefact free output with a stack of multiple nonlinear layers. However, looking only at the corrupted image data $x$ this can also be seen as a superposition of the ground truth and the artefact. In other words the artefact is simply the residual between the input and output $\Phi(x) = \Psi(x) - x$. Due to strong similarities in the artefact structures of different images, it is more reasonable to force the network to approximate the residual function $\Phi(x)$ and learn the artefact. By reformulating the problem, the initially desired mapping can then be obtained through $\Psi(x) = \Phi(x) + x$.

\subsection{Network Architecture}
\begin{figure}
\centering
\includegraphics[width=0.35\textwidth]{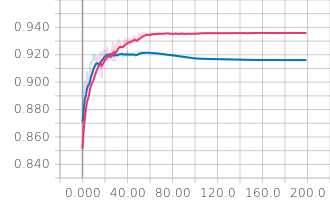}
\caption{Average accuracy on the validation set over the epochs while training on 1D Varden sampling - Pink shows the model with dropout (proposed model) and blue shows without dropout.}
\label{DOvsNoDO}
\end{figure}

\begin{figure*}
\centering
\includegraphics[width=0.95\textwidth]{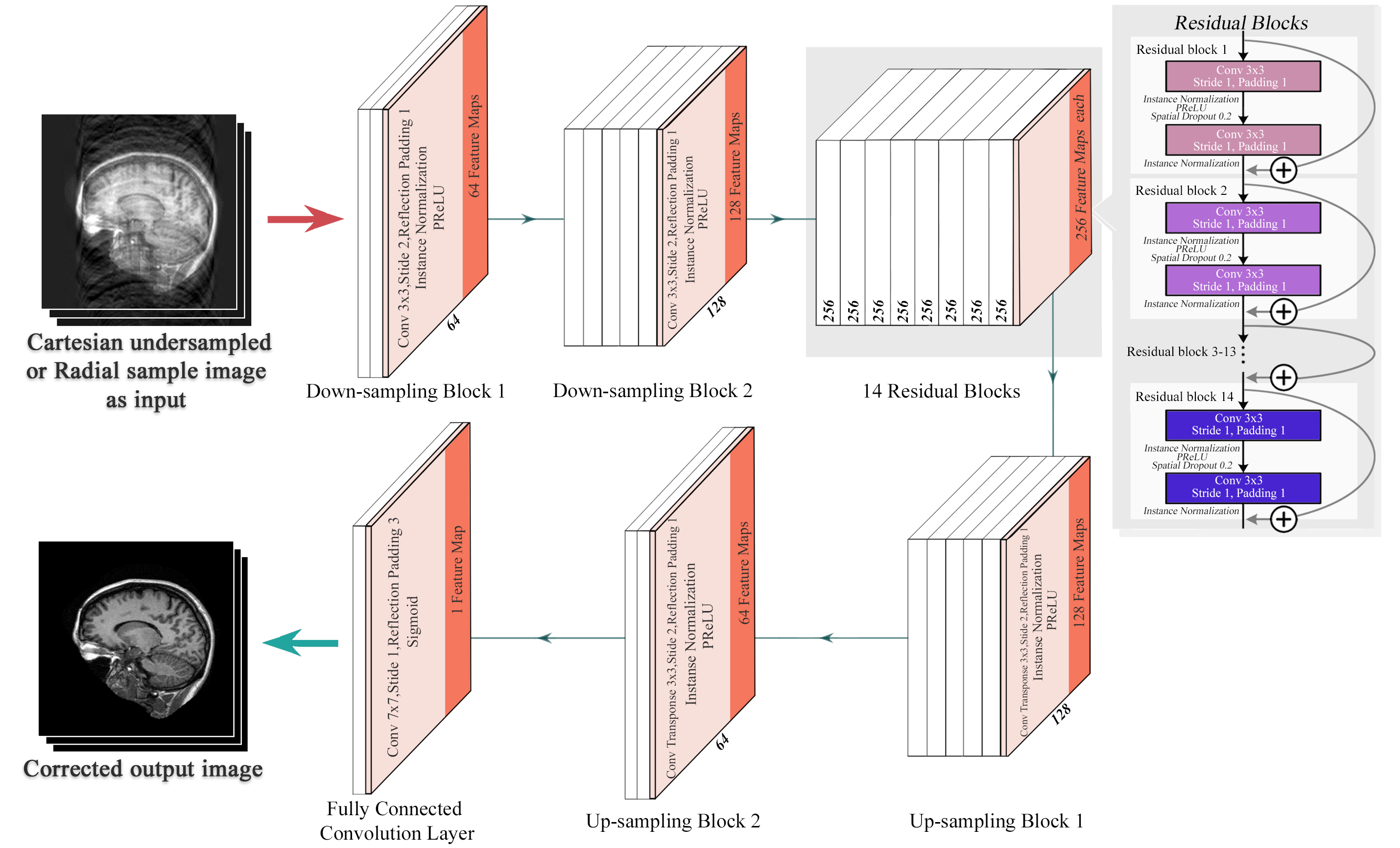}
\caption{ReconResNet: Architecture of the proposed network backbone}
\label{netarch}
\end{figure*}

The proposed network backbone, ReconResNet, has been created using the residual learning approach - ResNet~\citep{He.2016}. This paper proposes a modified regularised version of the Residual Block, by adding a Spatial Dropout~\citep{Tompson.16112014}, to randomly drop or zero out feature maps, between the two convolution layers of the Residual Block; this results in better accuracy on the validation set as shown in Fig.~\ref{DOvsNoDO}. In this model, first, the input is down-sampled with two down-sampling blocks, each decreases the input size by half in all dimensions, while increasing the number of feature maps by two (starting with 64). Next, the network contains 14 modified Residual Blocks. The input to each residual block is added to its output and forwarded to the next residual block – this is the reason behind the nomenclature – residual learning, as residue from the input is forwarded with the output to the next layer. After the residual blocks, the network contains two up-sampling blocks, each doubles the input size and reduces the number of feature maps by half, to obtain the original image size back. Finally, a fully-connected convolution layer is added, followed by Sigmoid as the final activation function. Furthermore, Parametric ReLU (PReLU) has been used as the activation function (except for the final activation), instead of the commonly used ReLU. The graphical representation of the network architecture is shown in Fig.~\ref{netarch}. 

\subsection{Data Consistency Step}
The data consistency step replaces the actual acquired undersampled data in the network's output. In this way, the final output is not entirely depended on the network. The network only helps to fill-in the data, which were ignored during the undersampled acquisition. 

\subsubsection{Cartesian sampling}
For undersampled Cartesian data, data consistency was performed following~\citet{ChangMinHyun.2018}. First, FFT was performed on the output image to get its corresponding k-space. An inverted sampling mask was applied to this to identify the k-space values, which were not acquired. Then the measured data were combined with the missing k-space data estimated by the network. Eventually, iFFT was applied to this combined k-space to obtain the final output.
 
\subsubsection{Radial sampling}
This paper introduces a data-consistency step for reconstruction of undersampled radial data. A sampling pattern was generated for a fully sampled image (referred as $\Omega_{FS}$), considered to have twice the number of spokes than its pixel resolution and a NUFFT~\citep{fessler2007nufft} object was created from it. Then using that object, a forward transform was performed on the output image of the network to obtain its fully sampled radial k-space. The measured spokes were then inserted into the output k-space. A density compensation function (DCF) was applied, followed by adjoint NUFFT using the same NUFFT object was performed to obtain the final output.

\subsection{Implementation} \label{sec:implement}
\begin{figure*}
\centering
\includegraphics[width=0.9\textwidth]{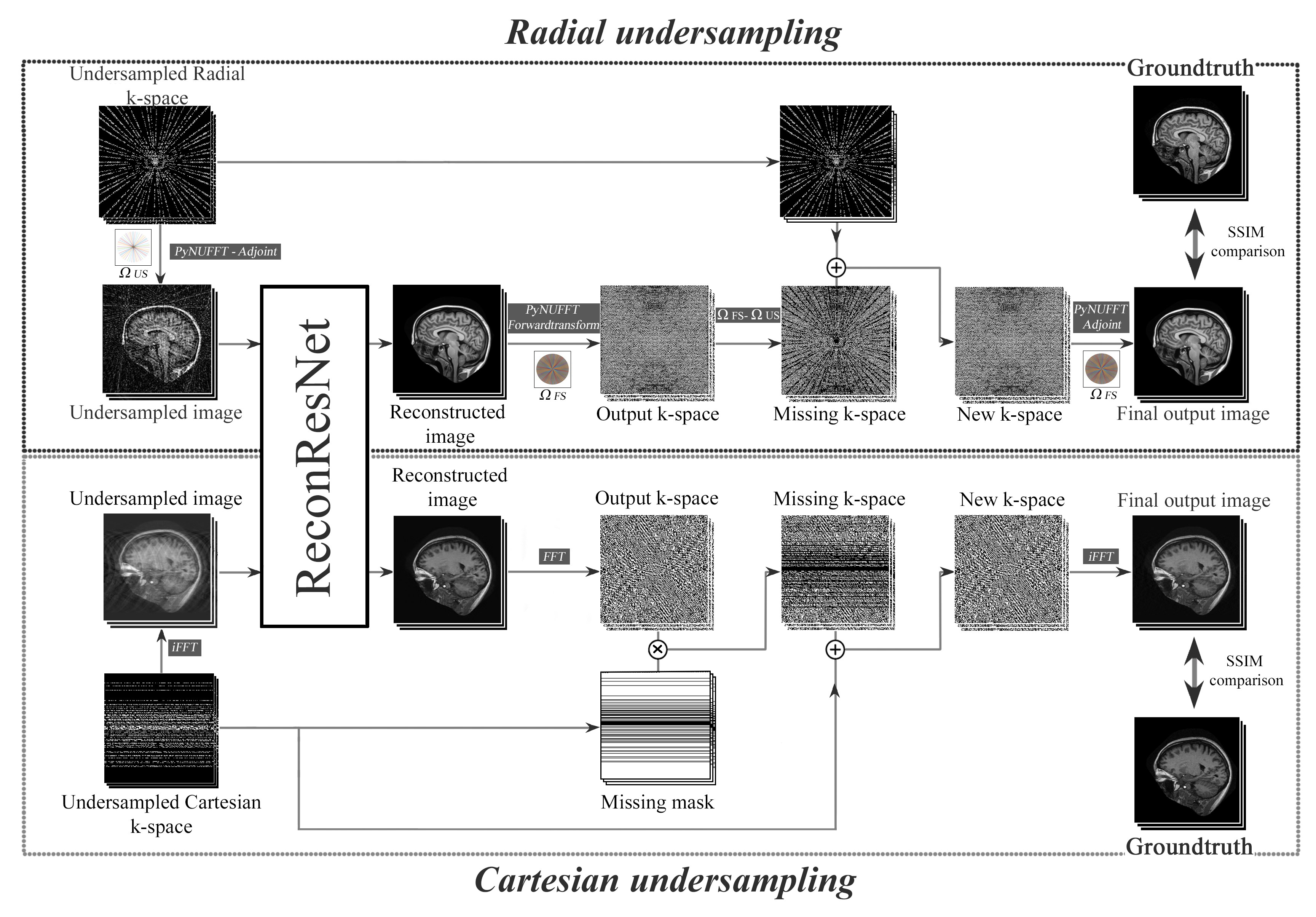}
\caption{Working Model} \label{workmodel}
\end{figure*}

Fig. \ref{workmodel} exhibits the working mechanism (including the data-flow) of the proposed framework, comprising a network backbone: ReconResNet, and data consistency steps for Cartesian and radial undersampling reconstructions. Only the network backbone is used during the training process. Whereas during inference, the complete framework is used. To train the ReconResNet, the loss has been calculated with the help of the Structural Similarity Index (SSIM)~\citep{Renieblas.2017} (range 0 to 1), where higher SSIM means closer image similarity. 
The SSIM value is calculated using the following equation:
\begin{equation}
    SSIM (x,y) = \frac{(2\mu_x\mu_y+C_1)(2\sigma_xy+C_2)}{(\mu_x^2+\mu_y^2+C_1)(\sigma_x^2+\sigma_y^2+C_2)}
    \label{eq:SSIM} 
\end{equation}
where $x$ and $y$ are the two images between which the structural similarity is to be calculated, $\mu_x, \mu_y, \sigma_x, \sigma_y$ and $\sigma_{xy}$ are the local means, standard deviations, and cross-covariance for images $x$ and $y$, respectively. $c_{1}=(k_{1}L)^{2}$ and $c_{2}=(k_{2}L)^{2}$, where $L$ is the dynamic range of the pixel-values, $k_{1}=0.01$ and $k_{2}=0.03$.
The negative of the SSIM value has been used as the loss value, and it was then be minimised using Adam Optimiser (Initial learning rate 0.0001, decayed by 10 after every 50 epochs; $\beta_1 = 0.9, \beta_2 = 0.999, \epsilon = 1e-09$), and was trained for 50 (for multi-slice models using all the slices), 100 (for multi-slice models, using only the central slices) or 200 epochs (for single-slice models) with a batch size of one. This network was implemented using PyTorch~\citep{NEURIPS2019_9015} and was trained using Nvidia GTX 980 and 1080Ti GPUs. The code of NCC1701 is available on GitHub~\footnote{NCC1701 on GitHub: \url{https://github.com/soumickmj/NCC1701}}.

The proposed ReconResNet is created using 2D convolutions, yet a 3D version using 3D convolutions, ReconResNet3D, was also created for comparison purposes. Due to the high memory requirement of such a model, ReconResNet3D was trained using an Nvidia Tesla V100 GPU. However, the number of starting feature maps had to be reduced to 32 (from 64 in the 2D version) due to GPU memory limitations.

\subsection{Dataset}
\subsubsection{Primary Dataset}
In this work, two different benchmark datasets were used. Firstly, OASIS-1~\citep{Marcus.2007}~dataset, which contains brain MRIs of 416 subjects acquired at 1.5T (details are in Table~\ref{tab:oasis_MR}), among which 150 were used for training and 100 were used for testing, selected randomly. 20\% of the training set were used as the validation set and the remaining 80\% as the actual training set. Only the first scan of each subject was used. Secondly, the IXI-dataset\footnote{\label{ixidataset}IXI Dataset: \url{https://brain-development.org/ixi-dataset/}.} was used, which contains nearly 600 Brain MRIs of normal healthy subjects of different contrasts (T1, T2, PD, MRA etc.) from three different hospitals and from two different field strengths (1.5T and 3T); T1 weighted volumes (see Table~\ref{tab:ixi_MR} for MR acquisition information) were used for training validation and testing, by randomly taking 100 volumes for each. Only healthy subjects were used for training to avoid any kind of pathology-related bias in the model. As the vast variability of the possible brain pathologies might be difficult to represent in the training set, the authors opted to train only on pathology-free datasets.

\begin{table}[!h]
\centering
\caption{OASIS-1 Dataset~\citep{Marcus.2007}~: MR acquisition information}
\label{tab:oasis_MR}
\begin{tabular}{@{}ll@{}}
\toprule
Field strength       & $1.5$T                       \\
Sequence             & MP-RAGE                    \\
Repetition time (TR) & $9.7\,ms$                  \\
Echo time (TE)       & $4.0\,ms$                  \\
Flip angle           & $10\degree$            \\
Inversion time (TI)  & $20\,ms$                   \\
Dead time (TD)       & $200\,ms$                  \\
Orientation          & Sagittal                   \\
Voxel size           & $1\times1\times1.25\,mm^3$ \\ \bottomrule
\end{tabular}
\end{table}

\begin{table}[!h]
\centering
\caption{IXI Dataset\cref{ixidataset}: MR acquisition information}
\label{tab:ixi_MR}
\resizebox{0.48\textwidth}{!}{%
\begin{tabular}{@{}lll@{}}
\toprule
                       & \begin{tabular}[c]{@{}l@{}}Guy's \\ Hospital\end{tabular} & \begin{tabular}[c]{@{}l@{}}Hammersmith \\ Hospital\end{tabular} \\ \midrule
Field strength         & $1.5$T                                                    & $3$T                                                              \\
Repetition time (TR)   & $9.81\,ms$                                                & $9.60\,ms$                                                      \\
Echo time (TE)         & $4.60\,ms$                                                & $4.60\,ms$                                                      \\
Flip angle             & $8\degree$                                            & $8\degree$                                                  \\
Orientation            & Sagittal                                                  & Sagittal                                                        \\
Voxel size (in $mm^3$) & $0.94\times0.94\times1.20$                                & $0.94\times0.94\times1.20$                                      \\ \bottomrule
\end{tabular}%
}
\end{table}

The proposed architecture works with 2D images. Therefore, preliminary tests were performed on the OASIS-1 dataset by taking slice number 50 out of all the MR volumes. Further experiments with 3D volumes were performed by creating a subset of each volume consisting of 30 slices (slice number 51 to 80). Each slice of the 3D volumes was sent individually to the framework as 2D slices, and the final outputs were later combined as a 3D volume. 

Tests on the IXI dataset were performed separately on 1.5T and 3T volumes, by using a subset of each volume containing 30 slices (slice number 61 to 90), to check the framework's performance at different field strengths. Further tests were performed taking all the slices of each 3T volume in two different manners: a 2D approach, where all the slices were supplied to ReconResNet separately and then combined to form 3D volumes; and a 3D approach, where the volumes were directly supplied to ReconResNet3D.

\subsubsection{Dataset for additional testing}
For training, only data from healthy subjects were used. To check the boundaries and validity of the approach in clinical settings, additional tests were performed on data of subjects with pathology. To test whether the proposed approach can reconstruct undersampled brain MRIs of Alzheimer patients, T1 weighted saggital MPRAGE images from the ADNI\footnote{Alzheimer's Disease Neuroimaging Initiative (ADNI) database:\url{https://adni.loni.usc.edu}.} dataset was used and for testing on brain MRIs with tumours, T1 contrast-enhanced images from the BraTS 2019\cite{menze2014multimodal} dataset was used. From ADNI, data of patients with Mild Cognitive Impairment (MCI), Significant Memory Concern (SMC), Alzheimer's disease (AD) were included. From both datasets, only T1 weighted images were taken, as the network was trained on a similar type of acquisition.

\subsection{Undersampling}
Images from all the datasets were treated as fully sampled images, for both Cartesian and radial sampled experiments. The datasets don't contain any raw MR data. Thus, simulated single-channel fully sampled raw data, as well as various undersampled datasets were artificially generated using MRUnder\footnote{MRUnder on Github: \url{https://github.com/soumickmj/MRUnder}}~\citep{soumick_chatterjee_2020_3901455} pipeline.

\subsubsection{Cartesian sampling}
\begin{figure}
\centering
\includegraphics[width=\columnwidth]{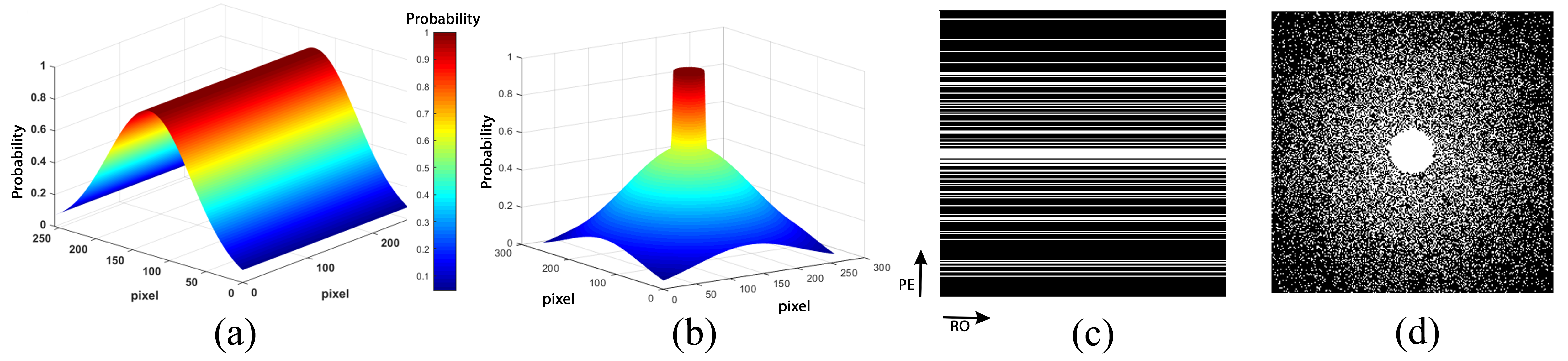}
\caption{From left to right (a) Probability density function for 1D Varden Mask (b) Probability density function for 2D Varden Mask (c) 1D Varden Mask and (d) 2D Varden Mask. All of them are for image size 256x256, taking 30\%of the k-space.}
\label{varden}
\end{figure}

Artificial undersampling of the Cartesian raw data was achieved using a k-space sampling pattern~(sampling~mask)~\citep{Lustig.2007}, which was generated by selecting fully sampled readout lines randomly in the phase encoding direction following a one-dimensional normal distribution (Fig.~\ref{varden}a) with the centre of the distribution matching the centre of the k-space (referred as 1D Varden). Also, a densely sampled centre consisting of eight lines was added. For the two-dimensional variable density mask~\citep{Lustig.2007} (referred as 2D Varden), randomly distributed k-space points were selected following a two-dimensional normal distribution~(Fig.~\ref{varden}b) with a densely sampled centre covering 2.5\% of the k-space. Furthermore, a uniform sampling mask was also generated by taking every n\textsuperscript{th} readout line from, where n equals to the step size, and a densely sampled centre consisting of nine lines was added.

The initial set of experiments was performed with three different Cartesian undersampling patterns. 1D and 2D Varden mask were generated by taking 30\% of the k-space, and a uniform sampling mask was generated with a step size of 4, resulting in approximately 28\% of the k-space. These patterns resulted in an acceleration factor of approximately 3.5. Further experiments were performed to check the performance of the framework while reconstructing higher acceleration factors, by taking 15\%, 10\% and 5\% of the k-space - resulting in approximate acceleration factors of 7, 10 and 20 respectively. 

\subsubsection{Radial sampling}
Artificially radially undersampled data were generated with the help of PyNUFFT~\citep{Lin.2018}. The sampling pattern $\Omega_{US}$ was defined with the golden angle of $111.25^\circ$~\citep{Feng.2014}~\citep{Block.2014} between the radial acquisitions (spokes). A forward NUFFT using a NUFFT object following $\Omega_{US}$ was performed on the fully sampled image data to obtain an undersampled radial k-space, then a density compensation function (DCF) was applied to compensate for the over-sampled centre. Finally, adjoint NUFFT~\citep{Lin.2018} was used to bring it back to the image space with radial undersampling artefacts. 

In this paper, two different levels of radial undersampling by taking 60 and 30 spokes were evaluated. Considering the fully-sampled data to have twice the number of spokes than its pixel resolution (here 256), the fully-sampled data had 512 spokes. Hence, k-space with 60 and 30 spokes correspond to approximate acceleration factors of 8.5 and 17 respectively.

\subsection{Evaluation criteria}
Results have been evaluated with the help of structural similarity index (SSIM, Eq.\ref{eq:SSIM}), mean-squared error (MSE, Eq.\ref{eq:mse}) and normalised root-mean-squared error (NRMSE, Eq.\ref{eq:nrmse}), where the normalisation factor was the averaged Euclidean norm of the fully sampled image. All the similarity measures were calculated by comparing the fully-sampled image with the under-sampled image (zero-filled k-space) or the output image. For quantitative evaluation of the reconstruction of images with Alzheimer’s disease, the Percentage Brain Volume Change (BVD, Eq.\ref{eq:pbvc}) was calculated from the normalised brain volume (NBV) obtained using FSL SIENAX~\citep{smith2004advances,woolrich2009bayesian,jenkinson2012fsl}. The BVD values were used to confirm brain volume preservation. Reconstruction performance of the models while reconstructing brain MRIs with tumours from the BraTS dataset was evaluated qualitatively by inspecting the tumour in the reconstruction.

\begin{equation}
\begin{aligned}
    {MSE} ={\frac {1}{n}}\sum _{i=1}^{n}(Y_{i}-{\hat {Y_{i}}})^{2}
\end{aligned}
\label{eq:mse}
\end{equation}

\begin{equation}
\begin{aligned}
    {NRMSE} = \frac{\sqrt{MSE}}{\sqrt{{\frac {1}{n}}\sum _{i=1}^{n}Y_{i}^{2}}}
\end{aligned}
\label{eq:nrmse}
\end{equation}

where $Y_{i}$ denotes the pixels of the ground-truth fully-sampled image, $\hat{Y_{i}}$ denotes the pixels of the undersampled image or the reconstruction (depending upon the comparison been performed) and $n$ denotes the number of pixels in the image.

\begin{equation}
\begin{aligned}
  {BVD}_{\left(fully-recon\right)}=\ \frac{\left|{NBV}_{fully}-{NBV}_{recon}\right|}{\overline{{NBV}_{all}}}\ast100 \\
  {BVD}_{\left(fully-under\right)}=\ \frac{\left|{NBV}_{fully}-{NBV}_{under}\right|}{\overline{{NBV}_{all}}}\ast100 
\end{aligned}
\label{eq:pbvc}
\end{equation}


\section{Results}\label{sec:Results}

\begin{figure*}[hbtp!]
\centering
\includegraphics[width=0.9\textwidth]{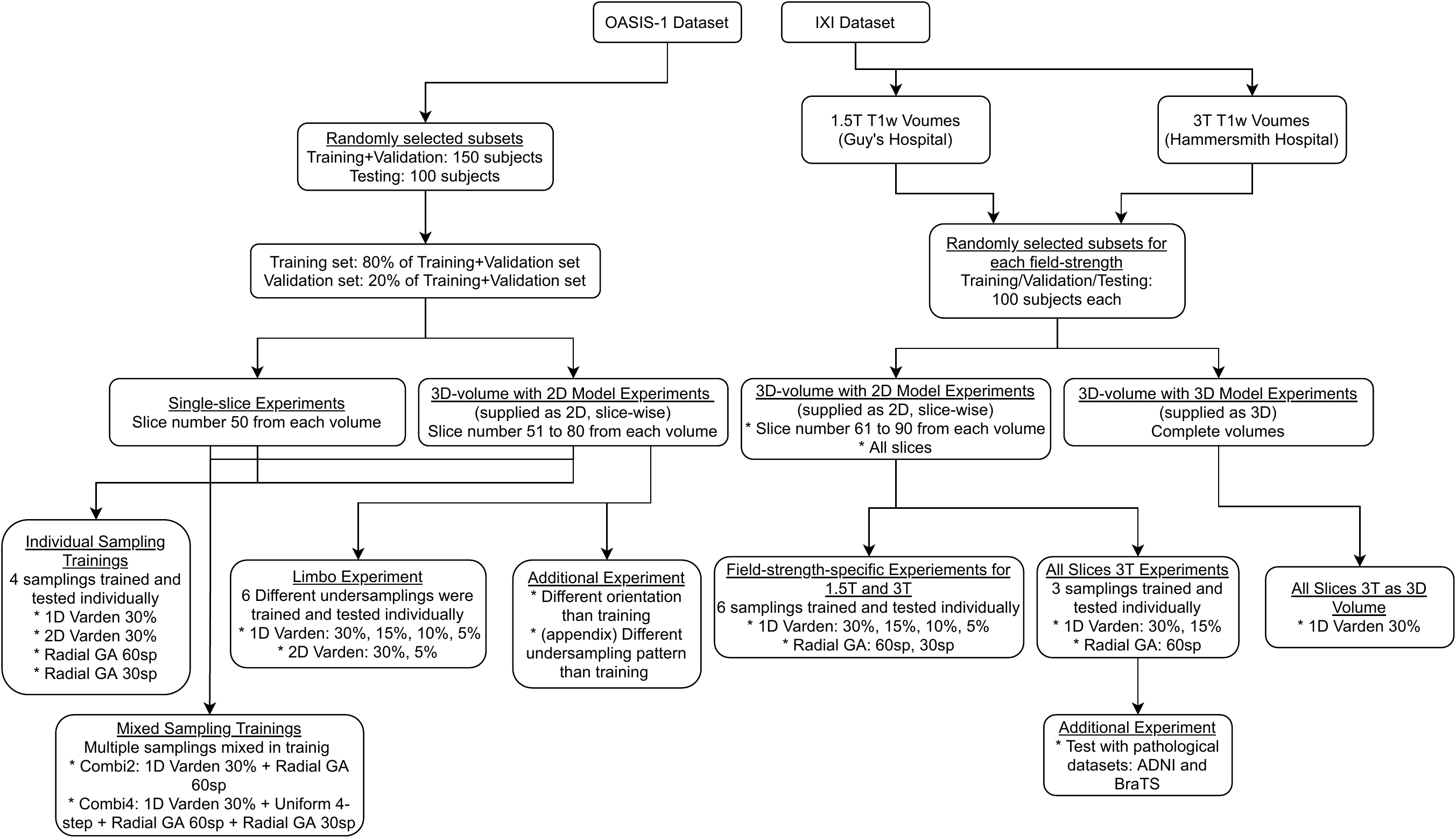}
\caption{Experiment Design} \label{fig:dataflow}
\end{figure*}

Various set of experiments were performed using both the training datasets. OASIS-1 dataset was used initially to compare the framework's performance when trained using one slice and then by training on multiple slices by choosing centre slices of the brain, framework's performance when multiple sampling patterns are trained together, and to evaluate the limits of the framework in terms of undersampling factor - known herewith as the "Limbo" experiment. Then experiments were performed with the IXI dataset for different undersampling patterns and different field strengths - by taking the central slices, by using all the slices, and also by supplying the 3D volumes to a 3D variant of the ReconResNet model. OASIS-1 dataset was also used to perform the additional experiments of the ability of the network while reconstructing orientations different than the training orientation and reconstructing different undersampling pattern than the training and IXI dataset trainings were used to evaluate pathological scenarios. The various experiments with the different datasets are summarised in Fig.~\ref{fig:dataflow}. 

Training OASIS-1 single slice models took trained on slice number 50 of each volume for 200 epochs took around 4 hours, training the OASIS and IXI multi-slice models (3D volumes, supplied as 2D slices to the models) trained on the central slices for 100 epochs took around 45-50 hours each, while the IXI models trained on the whole volumes (as 2D slices) for 50 epochs took around 95-100 hours each. The reconstruction time using this framework for the Cartesian undersampling patterns took only a fraction of a second for each slice, while the radial undersampling patterns took around a second. It is worth mentioning that the timings may differ depending upon the system configuration (GPU, CPU, etc.) and the current workload. Finally, before going to the results, it is worth mentioning that the training and validation curves were monitored to detect any possible overfitting, and none of the trainings showed any overfitting behaviour. Examples of training and validation curves have been provided in the Appendix.

\subsection{OASIS-1 dataset}
\label{sec:res_oasis}
\subsubsection{Various undersampling patterns trained independently}
\begin{figure*}
\centering
\includegraphics[width=\textwidth]{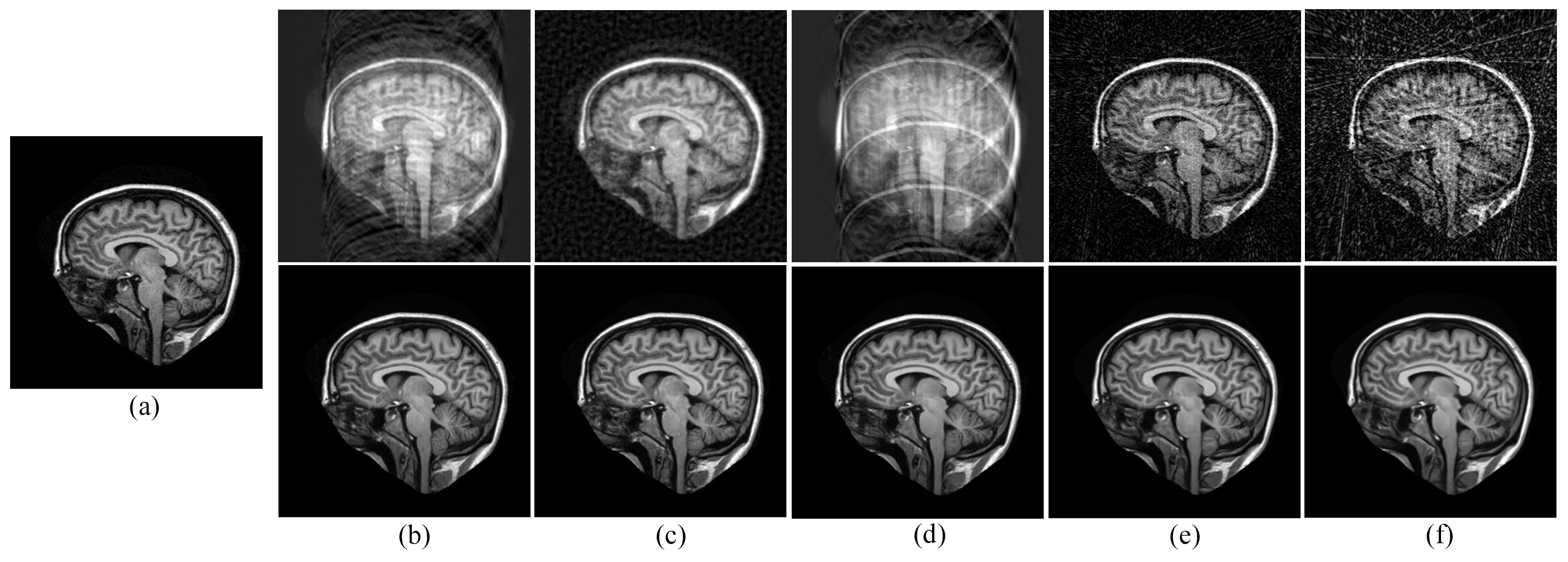}
\caption{Network's example output (OASIS-1 dataset). (a) Fully sampled. From b to f - top undersampled, bottom output of the framework. Undersampling patterns: (b) Cartesian 1D Varden 30\%, (c) Cartesian 2D Varden 30\%, (d) Cartesian Uniform sampling step size 4, (e) Radial 60 spokes, and (f) Radial 30 spokes.}
\label{fig:output}
\end{figure*}

For the first series of tests, various Cartesian and radial sampling patterns were trained and tested individually (results in Table~\ref{tab:tab1_OASIS_TrainSep_Cart}~and~\ref{tab:tab1_OASIS_TrainSep_Rad}). Two different types of trainings were performed: single slice training on slice 50 and trainings with 3D volumes (slice 51 to 80), supplied as 2D slices. While reconstructing 1D Varden 30\%, 2D Varden 30\%, Radial 60 spokes and Radial 30 spokes, the models yielded 0.986$\pm$0.005, 0.986$\pm$0.003, 0.981$\pm$0.004 and 0.947$\pm$0.011 SSIMs respectively. An example set of outputs for different sampling patterns has been shown in Fig.~\ref{fig:output} for qualitative evaluation.

\begin{table*}[]
\centering
\caption{Result of Cartesian undersampling patterns, while being trained separately}
\label{tab:tab1_OASIS_TrainSep_Cart}
\resizebox{\textwidth}{!}{%
\begin{tabular}{|c|c|c|c|c|c|c|c|}
\hline
\multicolumn{2}{|c|}{\multirow{2}{*}{}} & \multicolumn{3}{c|}{\textbf{1D Varden 30\%}}  & \multicolumn{3}{c|}{\textbf{2D Varden 30\%}}  \\ \cline{3-8} 
\multicolumn{2}{|c|}{}                  & \textit{SSIM} & \textit{MSE} & \textit{NRMSE} & \textit{SSIM} & \textit{MSE} & \textit{NRMSE} \\ \hline
\textbf{Single Slice} & \textit{Under}  & 0.336$\pm$0.019 & 0.0235    0.0036 & 0.80 $\pm$0.10 & 0.410$\pm$0.022 & 0.0103    0.0020 & 0.53 $\pm$0.07 \\
(Slice 50)            & \textit{Output} & 0.966$\pm$0.006 & 0.0006    0.0002 & 0.12 $\pm$0.02 & 0.971$\pm$0.003 & 0.0005    0.0002 & 0.11 $\pm$0.03 \\ \hline
\textbf{3D Volume}    & \textit{Under}  & 0.338$\pm$0.020 & 0.0234    0.0034 & 0.88 $\pm$0.12 & 0.427$\pm$0.022 & 0.0102    0.0019 & 0.58 $\pm$0.07 \\
(Slice 51-80)         & \textit{Output} & 0.986$\pm$0.005 & 0.0003    0.0002 & 0.09 $\pm$0.04 & 0.986$\pm$0.003 & 0.0003    0.0002 & 0.10 $\pm$0.03 \\ \hline
\end{tabular}%
}
\end{table*}

\begin{table*}[]
\centering
\caption{Result of Radial undersampling patterns, while being trained separately}
\label{tab:tab1_OASIS_TrainSep_Rad}
\resizebox{\textwidth}{!}{%
\begin{tabular}{|c|c|c|c|c|c|c|c|}
\hline
\multicolumn{2}{|c|}{\multirow{2}{*}{}} & \multicolumn{3}{c|}{\textbf{Radial 60sp}}     & \multicolumn{3}{c|}{\textbf{Radial 30sp}}     \\ \cline{3-8} 
\multicolumn{2}{|c|}{}                  & \textit{SSIM} & \textit{MSE} & \textit{NRMSE} & \textit{SSIM} & \textit{MSE} & \textit{NRMSE} \\ \hline
\textbf{Single Slice} & \textit{Under}  & 0.513$\pm$0.026 & 0.0024    0.0004 & 0.25 $\pm$0.02 & 0.375$\pm$0.019 & 0.0067    0.0010 & 0.42 $\pm$0.02 \\
(Slice 50)            & \textit{Output} & 0.960$\pm$0.007 & 0.0006    0.0002 & 0.12 $\pm$0.02 & 0.927$\pm$0.013 & 0.0010    0.0003 & 0.16 $\pm$0.02 \\ \hline
\textbf{3D Volume}    & \textit{Under}  & 0.539$\pm$0.024 & 0.0022    0.0003 & 0.27 $\pm$0.02 & 0.413$\pm$0.017 & 0.0061    0.0008 & 0.45 $\pm$0.03 \\
(Slice 51-80)         & \textit{Output} & 0.981$\pm$0.004 & 0.0004    0.0002 & 0.11 $\pm$0.03 & 0.947$\pm$0.011 & 0.0010    0.0002 & 0.18 $\pm$0.03 \\ \hline
\end{tabular}%
}
\end{table*}

\subsubsection{Cartesian and Radial undersampling trained together}
For testing the robustness of the framework further, different sampling patterns were combined together while training. For the first experiment, each image in the training and test set were undersampled using one Cartesian (1D Varden 30\%) and one radial (60 spokes golden-angle) sampling pattern (referred as Combi2, results in Table~\ref{tab:tab2_OASIS_Combo2}). For the second experiment, each image was undersampled using four different sampling patterns, two Cartesian (30\%~1D~Varden and uniform sampling with step size 4) and two radial (30 and 60 spokes golden-angle) (referred as Combi4, results in Table~\ref{tab:tab3_OASIS_Combo4}). 

In Combi2, a decrease of 0.20\% in SSIM can be observed for 1D Varden 30\% (0.984$\pm$0.004 for Combi2, 0.986$\pm$0.005 for independent) but an increase of 0.10\% for Radial 60 spokes (0.982$\pm$0.004 and 0.981$\pm$0.004). In Combi4, decrease of 0.81\% and 0.10\% can be observed for 1D Varden 30\% and Radial 60 spokes respectively, but an increase of 1.58\% for Radial 30 spokes (0.962$\pm$0.012 and 0.947$\pm$0.011). 

Further experiments were performed to test the capabilities of the model while reconstructing different undersampling patterns than the training pattern, are discussed in the Appendix. 

\begin{table*}[]
\centering
\caption{Result of Cartesian 1D variable density mask and Radial 60 spokes, while being trained together}
\label{tab:tab2_OASIS_Combo2}
\resizebox{\textwidth}{!}{%
\begin{tabular}{|c|c|c|c|c|c|c|c|}
\hline
\multicolumn{2}{|c|}{\multirow{2}{*}{}} & \multicolumn{3}{c|}{\textbf{1D Varden 30\%}}  & \multicolumn{3}{c|}{\textbf{Radial 60sp}}     \\ \cline{3-8} 
\multicolumn{2}{|c|}{}                  & \textit{SSIM} & \textit{MSE} & \textit{NRMSE} & \textit{SSIM} & \textit{MSE} & \textit{NRMSE} \\ \hline
\textbf{Single Slice} & \textit{Under}  & 0.336$\pm$0.019 & 0.0235    0.0036 & 0.80 $\pm$0.10 & 0.513$\pm$0.026 & 0.0024    0.0004 & 0.25 $\pm$0.02 \\
(Slice 50)            & \textit{Output} & 0.964$\pm$0.006 & 0.0006    0.0003 & 0.13 $\pm$0.03 & 0.963$\pm$0.007 & 0.0005    0.0002 & 0.12 $\pm$0.03 \\ \hline
\textbf{3D Volume}    & \textit{Under}  & 0.338$\pm$0.020 & 0.0234    0.0034 & 0.88 $\pm$0.12 & 0.540$\pm$0.024 & 0.0022    0.0003 & 0.27 $\pm$0.02 \\
(Slice 51-80)         & \textit{Output} & 0.984$\pm$0.004 & 0.0003    0.0002 & 0.10 $\pm$0.03 & 0.982$\pm$0.004 & 0.0003    0.0002 & 0.10 $\pm$0.03 \\ \hline
\end{tabular}%
}
\end{table*}

\begin{table*}[]
\centering
\caption{Result of Cartesian 1D variable density sampling, Uniform sampling and Radial 60, 30 spokes, while being trainedall four of the sampling pattern together}
\label{tab:tab3_OASIS_Combo4}
\resizebox{\textwidth}{!}{%
\begin{tabular}{|cc|cccc|c|c|c|c}
\hline
\multicolumn{1}{|l}{} &
  \multicolumn{1}{l|}{} &
  \multicolumn{4}{c}{\textbf{Cartesian}} &
  \multicolumn{4}{c|}{\textbf{Radial}} \\ \cline{3-10} 
 &
   &
  \multicolumn{2}{c|}{\textbf{1D Varden 30\%}} &
  \multicolumn{2}{c|}{\textbf{Uniform 4-step}} &
  \multicolumn{2}{c|}{\textbf{60 spokes}} &
  \multicolumn{2}{c|}{\textbf{30 spokes}} \\ \cline{3-10} 
 &
   &
  \multicolumn{1}{c|}{\textit{SSIM}} &
  \multicolumn{1}{c|}{\textit{NRMSE}} &
  \multicolumn{1}{c|}{\textit{SSIM}} &
  \textit{NRMSE} &
  \textit{SSIM} &
  \textit{NRMSE} &
  \textit{SSIM} &
  \multicolumn{1}{c|}{\textit{NRMSE}} \\ \hline
\multicolumn{1}{|c|}{\textbf{Single Slice}} &
  \textit{Under} &
  \multicolumn{1}{c|}{0.336$\pm$0.019} &
  \multicolumn{1}{c|}{0.80 $\pm$0.10} &
  \multicolumn{1}{c|}{0.282$\pm$0.018} &
  0.93 $\pm$0.12 &
  0.513$\pm$0.026 &
  0.25 $\pm$0.02 &
  0.375$\pm$0.019 &
  \multicolumn{1}{c|}{0.43 $\pm$0.02} \\
\multicolumn{1}{|c|}{(Slice 50)} &
  \textit{Output} &
  \multicolumn{1}{c|}{0.957$\pm$0.008} &
  \multicolumn{1}{c|}{0.14 $\pm$0.02} &
  \multicolumn{1}{c|}{0.947$\pm$0.012} &
  0.15 $\pm$0.02 &
  0.961$\pm$0.008 &
  0.12 $\pm$0.02 &
  0.935$\pm$0.011 &
  \multicolumn{1}{c|}{0.15 $\pm$0.02} \\ \hline
\multicolumn{1}{|c|}{\textbf{3D Volume}} &
  \textit{Under} &
  \multicolumn{1}{c|}{0.345$\pm$0.020} &
  \multicolumn{1}{c|}{0.88 $\pm$0.12} &
  \multicolumn{1}{c|}{0.289$\pm$0.018} &
  1.02 $\pm$0.14 &
  0.540$\pm$0.024 &
  0.27 $\pm$0.02 &
  0.413$\pm$0.018 &
  \multicolumn{1}{c|}{0.45 $\pm$0.03} \\
\multicolumn{1}{|c|}{(Slice 51-80)} &
  \textit{Output} &
  \multicolumn{1}{c|}{0.978$\pm$0.005} &
  \multicolumn{1}{c|}{0.12 $\pm$0.03} &
  \multicolumn{1}{c|}{0.968$\pm$0.010} &
  0.14 $\pm$0.03 &
  0.980$\pm$0.005 &
  0.11 $\pm$0.03 &
  0.962$\pm$0.012 &
  \multicolumn{1}{c|}{0.15 $\pm$0.03} \\ \hline
\end{tabular}%
}
\end{table*}

\subsubsection{Limbo: How low can it go?}
\label{sec:oasis_limbo}
\begin{figure}
    \centering
    \includegraphics[width=0.48\textwidth]{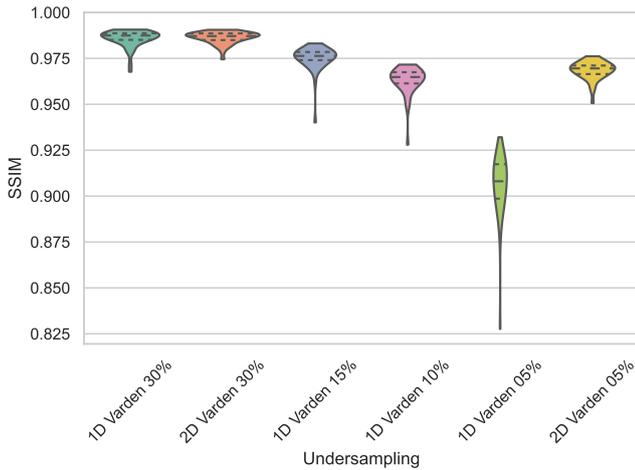}
    \caption{OASIS Limbo}
    \label{fig:tab4_OASIS_Limbo}
\end{figure}

The limitations of the network in terms of acceleration factors were evaluated using six different Cartesian undersampling patterns. Four levels of 1D Varden (30\%, 15\%, 10\% and 5\% of the k-space) and two levels of 2D Varden (30\% and 5\%) undersampling were used in this OASIS Limbo experiment. The quantitative results are been portrayed with the help of violin plots in Fig. \ref{fig:tab4_OASIS_Limbo}. For 30\% of the k-space, 1D Varden and 2D Varden sampling resulted in similar SSIM values: 0.986$\pm$0.005 and 0.986$\pm$0.003 respectively. 15\% and 10\% 1D Varden resulted in 0.975$\pm$0.007 and 0.963$\pm$0.008 respectively. When it comes to the highest acceleration factor evaluated in this research, 5\% of the k-space, 2D Varden resulted in 0.968$\pm$0.005 SSIM, but 1D Varden resulted in 0.906$\pm$0.017. Looking at the distribution in Fig. \ref{fig:tab4_OASIS_Limbo}, it can be said that the 1D Varden 5\% can not yield reliable results (while still resulting in aesthetically good but inaccurate results, further discussed in the Appendix) and other of the undersampling patterns resulted in similar performances. 

\subsection{IXI dataset}
\begin{figure}
    \centering
    \includegraphics[width=0.48\textwidth]{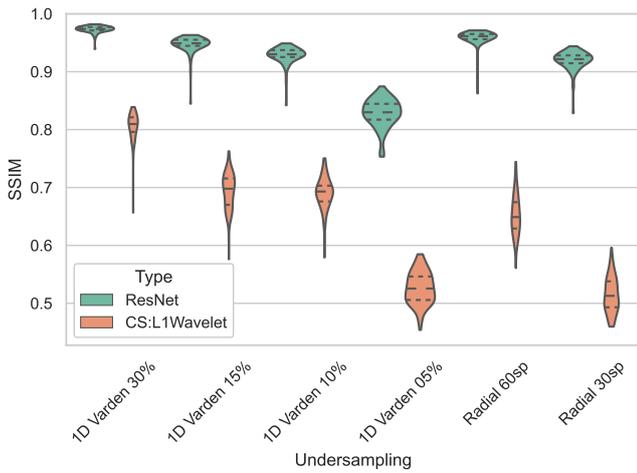}
    \caption{IXI 1.5T}
    \label{fig:tab6_IXIGuys_ResNetVCS}
\end{figure}

\begin{figure}
    \centering
    \includegraphics[width=0.48\textwidth]{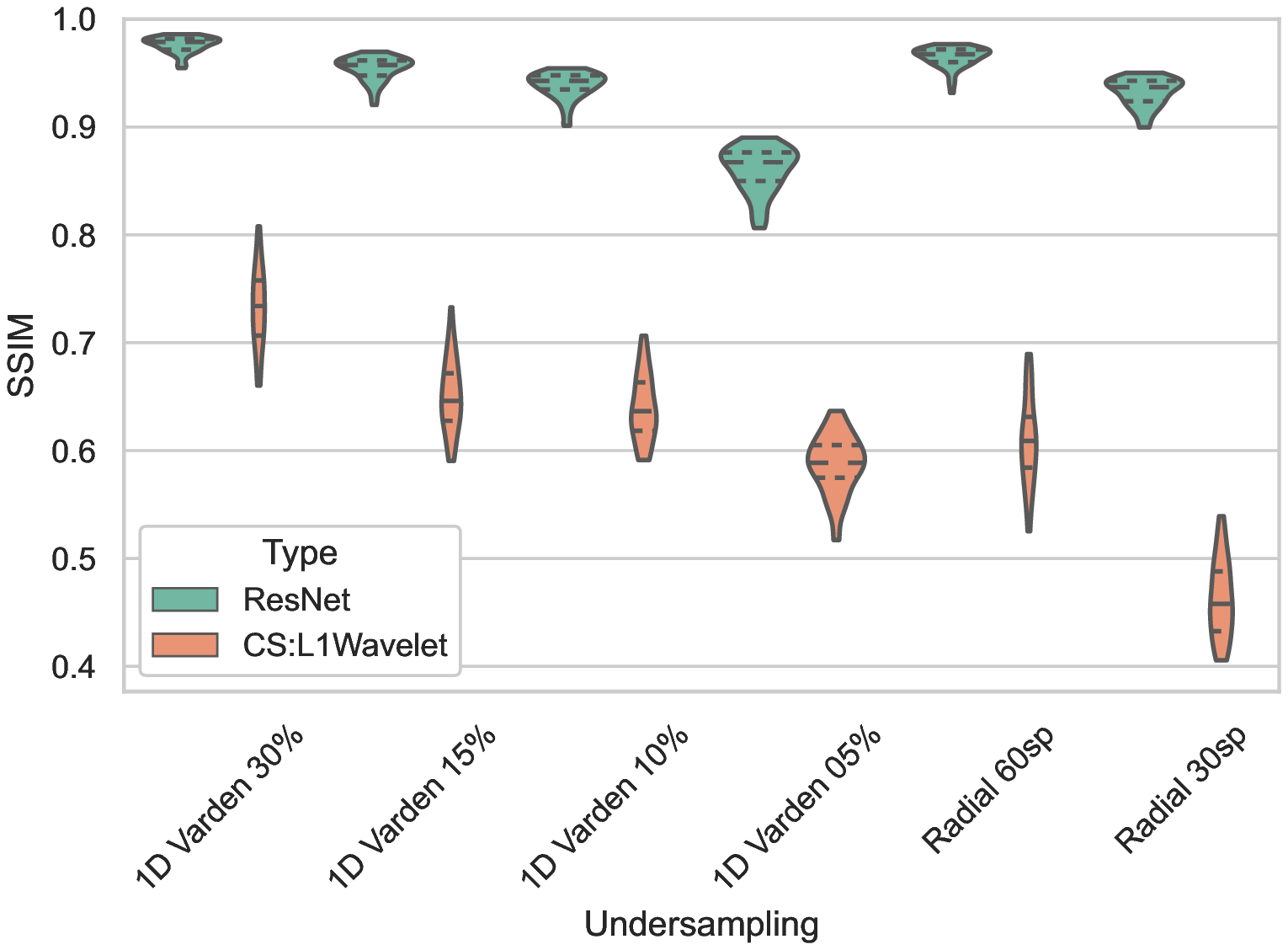}
    \caption{IXI 3T}
    \label{fig:tab5_IXIHH_ResNetVCS}
\end{figure}

The experiments with the IXI dataset were separated based on the field strengths (1.5T and 3T). Furthermore, the performance of the proposed framework was compared against compressed sensing (CS) reconstruction. Initially, two different CS reconstruction algorithm were evaluated for 1D Varden 30\% (3T IXI) - L1 Wavelet regularised reconstruction~\citep{Lustig.2007} and total variation (TV) regularised reconstruction~\citep{block2007undersampled}, resulted in 0.732$\pm$0.036 and 0.701 $\pm$0.04 SSIM values respectively. Given that L1 wavelet resulted in higher SSIM values in this experiment and since it has been used successfully in various studies~\citep{he2020dynamic,ueda2021compressed}, L1 wavelet was used for all the remaining experiments. Both L1 and TV reconstructions were performed using SigPy~\citep{ong2019sigpy} for 100 iterations with the regularisation parameter $\lambda=0.01$. IXI dataset contains already coil-combined images, and to assist the CS algorithms coil images were simulated using four-channel birdcage coil profile before undersampling. These coil-simulated images were then supplied to the CS algorithms. However, for RecoResNet the original coil-combined images were used. Four different levels of 1D Varden sampling (30\%, 15\%, 10\% and 5\% of the k-space) and two different levels of radial sampling (60 and 30 spokes) were evaluated on T1w MRIs acquired at 3T and 1.5T scanners.
These set of experiments were performed using a subset of the slices (slice 61 to 90) from each volume. The quantitative results are shown in Tab. \ref{tab:tab5n6_IXI_1p5n3}. Fig. \ref{fig:tab6_IXIGuys_ResNetVCS} and Fig. \ref{fig:tab5_IXIHH_ResNetVCS} portray the distribution of the resultant SSIM values. It can be observed that the proposed method was able to reconstruct both 1.5T and 3T images significantly better than the L1 wavelet regularised reconstruction for all the undersampling patterns.  It can be further said that the proposed network was able to reconstruct with high SSIM values even for higher undersampling factors, for 1D Varden Cartesian sampling acceleration factor 10 resulted in 0.939$\pm$0.012 and for radial sampling acceleration factor 17 resulted in 0.932$\pm$0.013 SSIM values for 3T IXI dataset. However, 1D Varden with acceleration factor 20 resulted in unreliable results (0.860$\pm$0.021 for 3T and 0.828$\pm$0.025 for 1.5T). This is in-line with the findings of the OASIS Limbo experiment discussed earlier.

\begin{table*}[]
\centering
\caption{Quantitative results while reconstructing different undersampling patterns, of 1.5T and 3T volumes from IXI dataset}
\label{tab:tab5n6_IXI_1p5n3}
\resizebox{0.90\textwidth}{!}{%
\begin{tabular}{|c|c|c|c|c|c|c|c|}
\hline
\multicolumn{2}{|c|}{\multirow{2}{*}{}} &
  \multicolumn{2}{c|}{\textbf{SSIM}} &
  \multicolumn{2}{c|}{\textbf{MSE}} &
  \multicolumn{2}{c|}{\textbf{NRMSE}} \\ \cline{3-8} 
\multicolumn{2}{|c|}{} &
  \multicolumn{1}{c|}{\textit{CS:L1Wavelet}} &
  \multicolumn{1}{c|}{\textit{ReconResNet}} &
  \multicolumn{1}{c|}{\textit{CS:L1Wavelet}} &
  \multicolumn{1}{c|}{\textit{ReconResNet}} &
  \multicolumn{1}{c|}{\textit{CS:L1Wavelet}} &
  \multicolumn{1}{c|}{\textit{ReconResNet}} \\ \hline
\textbf{1D Varden 30\%} &
  \textit{1.5T} &
  \multicolumn{1}{c|}{0.806$\pm$0.024} &
  \multicolumn{1}{c|}{0.973$\pm$0.006} &
  \multicolumn{1}{c|}{0.0066    0.0026} &
  \multicolumn{1}{c|}{0.0006    0.0002} &
  \multicolumn{1}{c|}{0.34 $\pm$0.08} &
  \multicolumn{1}{c|}{0.10 $\pm$0.02} \\
(AF: 3.5) &
  \textit{3T} &
  \multicolumn{1}{c|}{0.732$\pm$0.036} &
  \multicolumn{1}{c|}{0.976$\pm$0.007} &
  \multicolumn{1}{c|}{0.0125    0.0037} &
  \multicolumn{1}{c|}{0.0004    0.0003} &
  \multicolumn{1}{c|}{0.62 $\pm$0.11} &
  \multicolumn{1}{c|}{0.11 $\pm$0.03} \\ \hline
\textbf{1D Varden 15\%} &
  \textit{1.5T} &
  \multicolumn{1}{c|}{0.694$\pm$0.031} &
  \multicolumn{1}{c|}{0.948$\pm$0.014} &
  \multicolumn{1}{c|}{0.0097    0.0031} &
  \multicolumn{1}{c|}{0.0012    0.0010} &
  \multicolumn{1}{c|}{0.42 $\pm$0.09} &
  \multicolumn{1}{c|}{0.14 $\pm$0.05} \\
(AF: 7) &
  \textit{3T} &
  \multicolumn{1}{c|}{\textit{0.651$\pm$0.033}} &
  \multicolumn{1}{c|}{0.954$\pm$0.011} &
  \multicolumn{1}{c|}{0.0171    0.0041} &
  \multicolumn{1}{c|}{0.0008    0.0004} &
  \multicolumn{1}{c|}{0.73 $\pm$0.11} &
  \multicolumn{1}{c|}{0.15 $\pm$0.03} \\ \hline
\textbf{1D Varden 10\%} &
  \textit{1.5T} &
  0.690$\pm$0.028 &
  0.929$\pm$0.014 &
  0.0099    0.0032 &
  0.0016    0.0008 &
  0.42 $\pm$0.09 &
  0.17 $\pm$0.04 \\
(AF: 10) &
  \textit{3T} &
  0.639$\pm$0.030 &
  0.939$\pm$0.012 &
  0.0181    0.0038 &
  0.0009    0.0005 &
  0.76 $\pm$0.11 &
  0.17 $\pm$0.03 \\ \hline
\textbf{1D Varden 5\%} &
  \textit{1.5T} &
  0.526$\pm$0.028 &
  0.828$\pm$0.025 &
  0.0112    0.0023 &
  0.0041    0.0012 &
  0.45 $\pm$0.06 &
  0.27 $\pm$0.05 \\
(AF: 20) &
  \textit{3T} &
  0.587$\pm$0.027 &
  0.860$\pm$0.021 &
  0.0142    0.0033 &
  0.0020    0.0006 &
  0.67 $\pm$0.10 &
  0.25 $\pm$0.03 \\ \hline
\textbf{Radial 60 spokes} &
  \textit{1.5T} &
  0.652$\pm$0.034 &
  0.959$\pm$0.012 &
  0.0061    0.0021 &
  0.0009    0.0008 &
  0.33 $\pm$0.07 &
  0.13 $\pm$0.05 \\
(AF: 8.5) &
  \textit{3T} &
  0.608$\pm$0.040 &
  0.965$\pm$0.009 &
  0.0119    0.0035 &
  0.0006    0.0004 &
  0.61 $\pm$0.11 &
  0.13 $\pm$0.03 \\ \hline
\textbf{Radial 30 spokes} &
  \textit{1.5T} &
  0.514$\pm$0.030 &
  0.920$\pm$0.016 &
  0.0080    0.0010 &
  0.0017    0.0009 &
  0.38 $\pm$0.05 &
  0.17 $\pm$0.04 \\
(AF: 17) &
  \textit{3T} &
  0.462$\pm$0.035 &
  0.932$\pm$0.013 &
  0.0119    0.0022 &
  0.0010    0.0004 &
  0.62 $\pm$0.11 &
  0.11 $\pm$0.03 \\ \hline
\end{tabular}%
}
\end{table*}

\subsubsection{Whole volume experiments}
\begin{figure}
    \centering
    \includegraphics[width=0.48\textwidth]{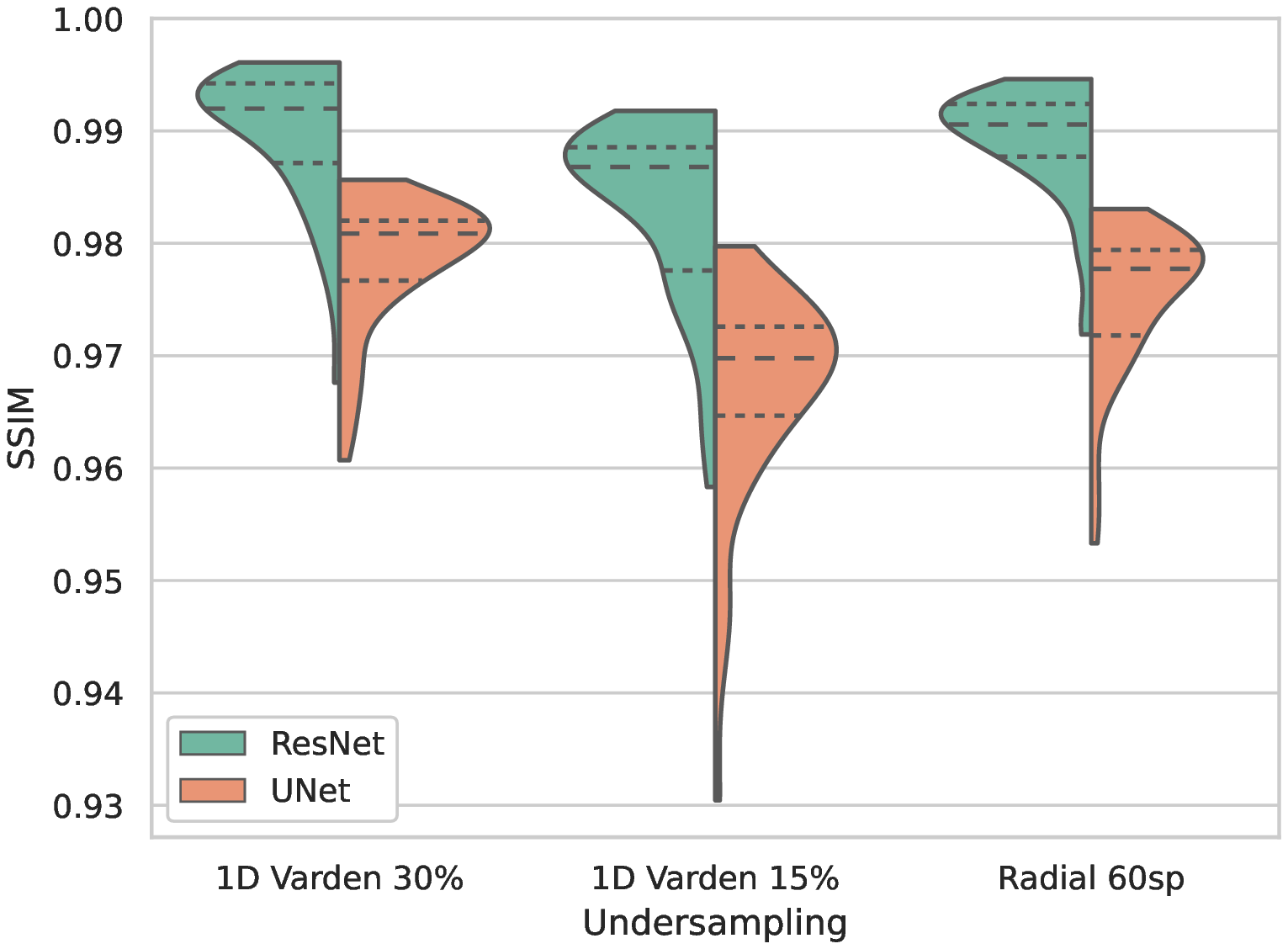}
    \caption{IXI 3T Full Vol}
    \label{fig:tab7_IXIHH_ResNet_FullVol}
\end{figure}

Further experiments were performed using the whole volumes for training and testing. These experiments were performed using the 3T volumes, by supplying them as 2D slices to the network, with three different undersampling patterns: 1D Varden 30\%, 1D Varden 15\% and Radial 60 spokes. The results of the proposed ReconResNet were compared against the U-Net \citep{ChangMinHyun.2018} results, and hypothesis testing was performed using independent two-sample t-test. The proposed model performed successfully while achieving statistically significant improvement over U-Net (p-values always less than $10^{-14}$), for all three sampling patterns and ReconResNet resulted in 0.990$\pm$0.006, 0.983$\pm$0.008 and 0.989$\pm$0.005 SSIM values respectively. Fig \ref{fig:tab7_IXIHH_ResNet_FullVol} shows the quantitative results with the help of violin plots.

\begin{figure}
    \centering
    \includegraphics[width=0.48\textwidth]{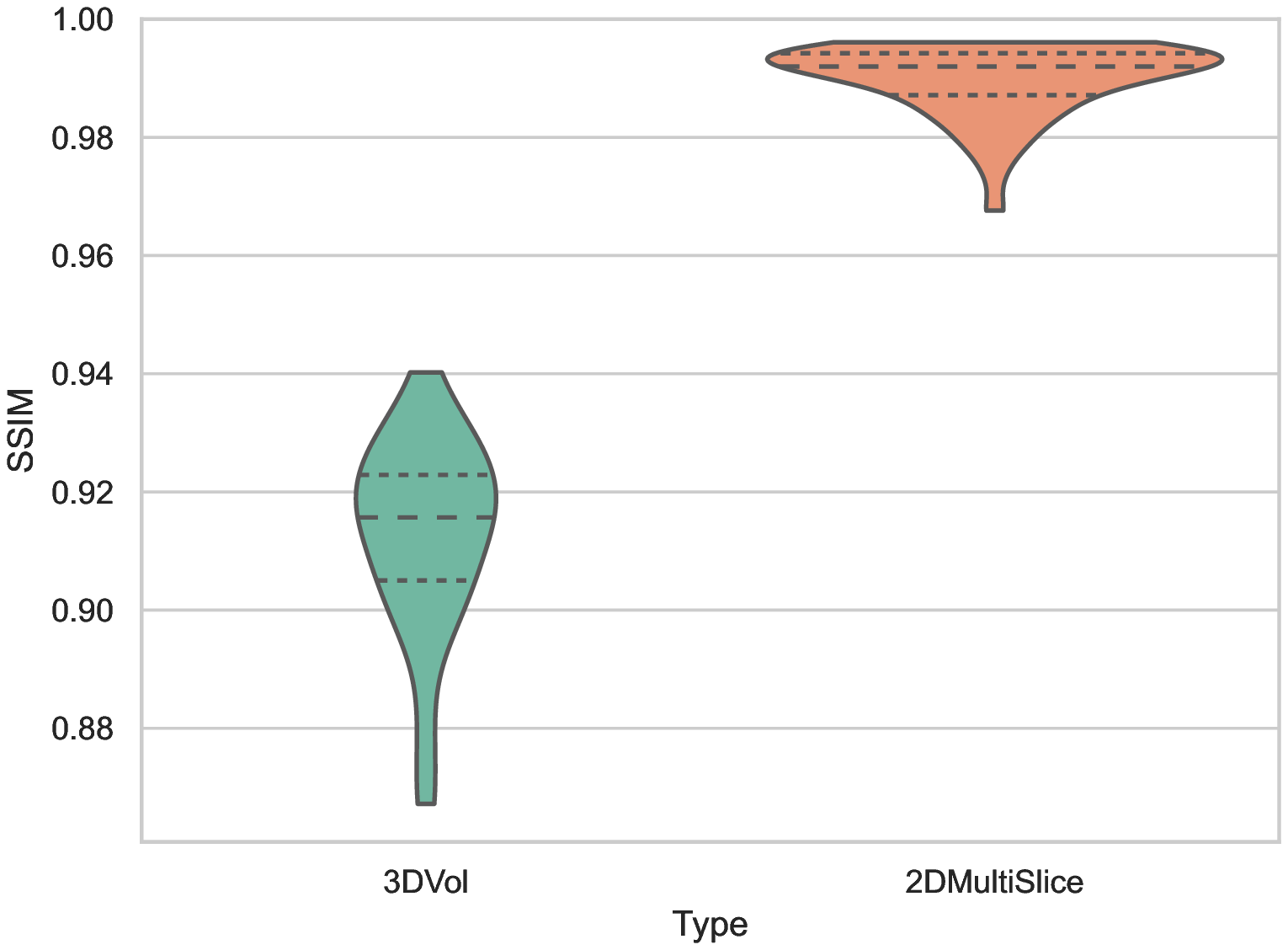}
    \caption{IXI 3T Full Vol - ReconResNet 3D vs 2D}
    \label{fig:tab7_IXIHH_ResNet_FullVol_3D}
\end{figure}

One additional experiment was performed by directly supplying the 3D 3T volumes undersampling using 1D Varden 30\% to the framework using ReconResNet3D as the backbone network. It was observed that the performance of the network was  7.88\% lower in terms of SSIM values than the 2D version, resulting in 0.912$\pm$0.017. Fig. \ref{fig:tab7_IXIHH_ResNet_FullVol_3D} shows the comparative results of both the models.
 
\subsection{Additional experiments}
\subsubsection{Reconstructing images in different orientations than the training}

\begin{figure*}
    \centering
    \includegraphics[width=\textwidth]{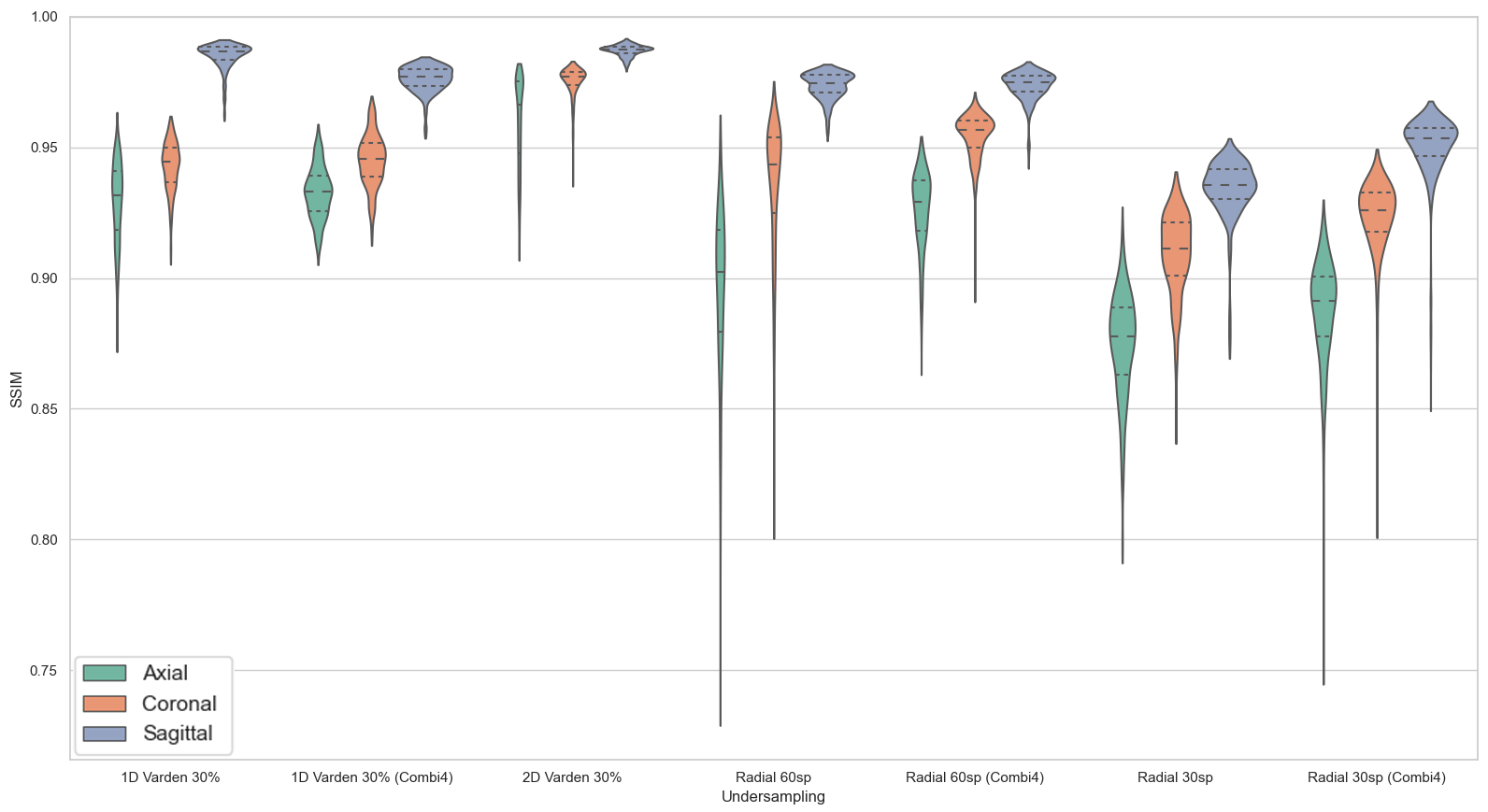}
    \caption{SSIM values for 2D Models Tested on Original orientation (Sagittal in blue) or Different Orientation (Coronal and Axial in orange and green respectively). In Combi4, the model was trained on Radial 60sp, Radial 30sp, Uniform4step and Varden 1D 30\%.}
    \label{fig:ssim_diff_orient}
\end{figure*}

\begin{figure*}
    \centering
    \includegraphics[width=\textwidth]{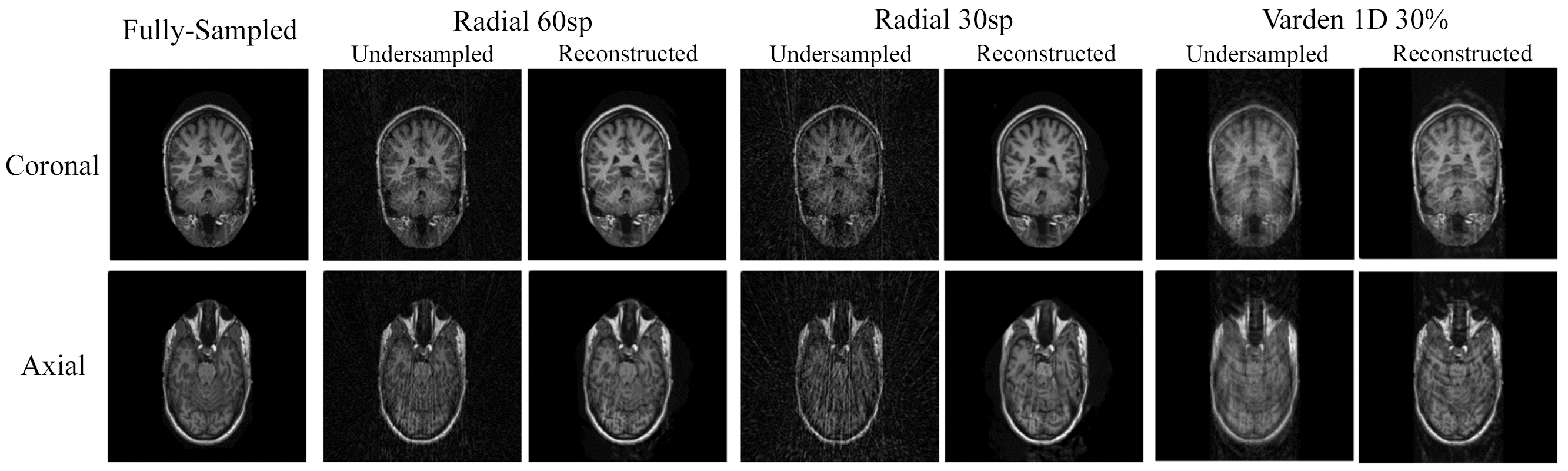}
    \caption{The top row is an example of the reconstruction of the coronal orientation and the axial orientation is shown in the bottom row: For Radial 60sp, 30sp and 1D Varden 30\%}
    \label{fig:img_diff_orient}
\end{figure*}

The models trained using OASIS-1 dataset were used in testing the performance of the network while reconstructing different orientations (coronal and axial) from the one used in the training (sagittal). As observed in Fig.\ref{fig:ssim_diff_orient}, testing the models using the coronal orientation yield higher SSIM values than the axial orientation. The models trained on a combination of undersampling patterns (Combi4) did better in reconstructing different orientations scoring higher SSIM values than the specific models trained on one pattern. As illustrated in Fig.\ref{fig:img_diff_orient} reconstructions of coronal orientations are better than reconstructions of axial orientations, where some artefacts are still present. It also shows a better reconstruction in case of radial undersampling patterns in comparison to Cartesian for both axial and coronal orientations. 

\subsubsection{Reconstructing brain MRIs with pathological abnormalities}
For testing the network’s applicability onto clinical data, MR volumes of Alzheimer's disease patients and patients with brain tumours were used from the ADNI and BraTS datasets respectively. 

\begin{figure*}
    \centering
    \includegraphics[width=\textwidth]{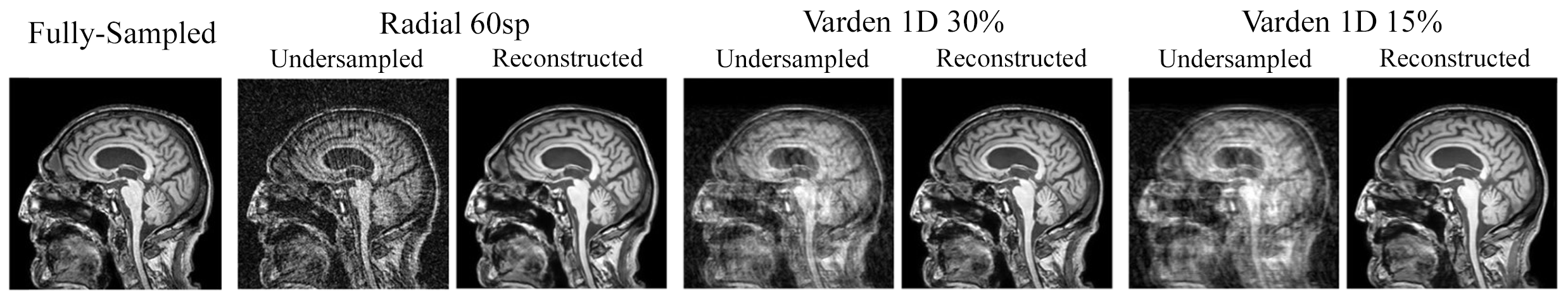}
    \caption{Examples of undersampled reconstruction of ADNI images: Radial 60sp, 1D Varden 30\% and 15\%}
    \label{fig:img_adni}
\end{figure*}

\begin{figure}
    \centering
    \includegraphics[width=0.49\textwidth]{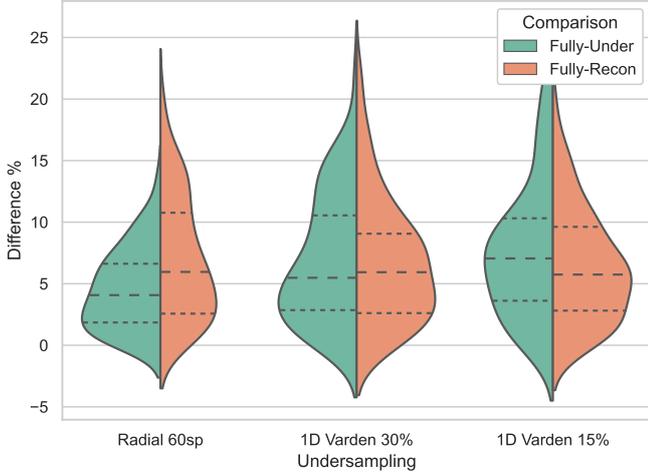}
    \caption{Paired comparison of brain-volumes (BVD) values for (fully-sampled-reconstructed) in green and (fully-sampled-under) in orange or each tested model. }
    \label{fig:pbvc_adni}
\end{figure}

For the ADNI dataset, the performance was measured quantitatively by measuring the difference in brain-volume (BVD, Eq.\ref{eq:pbvc}), to evaluate the framework's ability in preserving the brain volume. Fig.\ref{fig:pbvc_adni} shows that the BVD values between fully-sampled-reconstructed and fully-sampled-undersampled for both radial and Cartesian undersampling patterns. Fig.\ref{fig:img_adni} shows the reconstructed output, which is free from artefacts and highly similar to the fully-sampled image.

\begin{figure}
    \centering
    \includegraphics[width=0.49\textwidth]{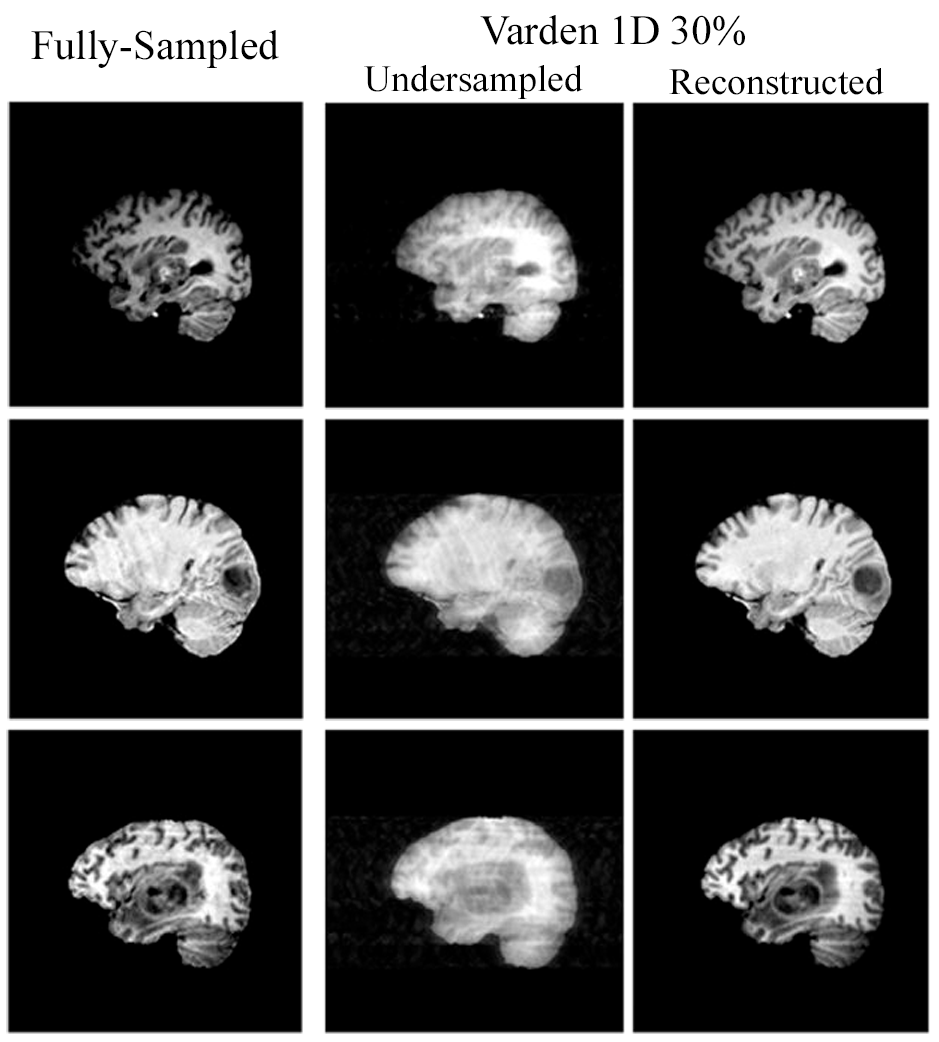}
    \caption{Examples of the BraTS 2019 dataset outcome tested on Varden 1D 30\% model. It can be seen that the tumours are preserved and most of the artefacts are gone.}
    \label{fig:img_brats}
\end{figure}

As for testing the clinical tumour data, the performance was evaluated using qualitative assessment. Fig.\ref{fig:img_brats} shows visual examples of the reconstructed images, where the tumour seems to be preserved and most of the artefacts are gone. 

\section{Discussion}\label{sec:Discussion}
The evaluation of the proposed framework for undersampled MRI reconstruction has proven its applicability based on different datasets. It can be observed that the framework seems to be robust against various undersampling patterns. When different sampling patterns were trained together, the accuracy is a little bit lower. For example, for 3D volumes undersampled using 1D Varden 30\%, when trained only on this sampling pattern resulted in 0.986$\pm$0.005, while being trained along with radial 60 spokes resulted in 0.984$\pm$0.004, and while being trained with three other sampling patterns resulted in 0.978$\pm$0.005 SSIM. For radial 60, the resultant SSIM values are similar for all three cases. For radial 30 spokes, the result improved when the training was performed in combination with other patterns (Combi4: 0.962$\pm$0.012, independent: 0.947$\pm$0.011). So, it can be said that the network can be trained with various types of sampling patterns all at the same time to make the trained network work with various sampling patterns. 

OASIS Limbo experiments have shown that the framework was able to reconstruct undersampling factors as high as 20 for 2D Varden sampling and 10 for 1D Varden sampling, yielded 0.968$\pm$0.005 and 0.963$\pm$0.008 respectively. However, the network struggled to reconstruct an acceleration factor of 20 for 1D Varden sampling (resultant SSIM: 0.906$\pm$0.017). With the help of IXI dataset, it was shown that the network was able to perform well for both 1.5T and 3T MRIs, for up to an acceleration factor of 17 for radial sampling and an acceleration factor of 10 for 1D Varden sampling, resulting in 0.932$\pm$0.013 and 0.939$\pm$0.012 respectively for 3T, 0.920$\pm$0.016 and 0.929$\pm$0.014 respectively for 1.5T. It was further observed that the proposed ReconResNet performed significantly better than L1-wavelet regularised compressed sensing reconstruction and U-Net.

While experimenting with the whole 3D volume, it was observed that the 2D version of the model trained on the slices separately worked superior to the ReconResNet3D working with the whole volume as 3D, resulting in 0.990$\pm$0.006 and 0.912$\pm$0.017 SSIM values respectively for 3T volumes undersampled with 1D Varden 30\%. It should be taken into account that these two models are not directly comparable as the number of starting features was cut by half (from 64 to 32) for the 3D model due to GPU memory limitations, even while using more powerful GPU than the 2D model (discussed in Sec. \ref{sec:implement}). Moreover, it is to be noted that the 3D model received significantly less number of data-points than the 2D model, albeit each data-point is larger: because the 3D model received the slices combined together as a volume, where the 2D model received them individually. This might also have an negative impact on the results of the 3D model. Given these results and also taking into account the GPU memory demand of ReconResNet3D, it can be said that the 2D model might sufficient to perform this task of undersampled MRI reconstruction. However, this demands further experimentation to reach a definitive conclusion. 

The experiments performed while reconstructing images in different orientations than the training orientation showed that the reconstruction of images in coronal orientation was more successful than in the axial orientation, while the network has been trained on sagittal images. One possible explanation is that sagittal and coronal orientations share more structural similarities (for example, the resemblance of the cerebellum and brain-stem) than sagittal and axial images. Besides, the eyes optical nerve structure in axial images is more pronounced and differs from sagittal and coronal orientation. As for the difference in the performance between the network trained on multiple patterns and on a specific one, one can argue that it is better for the network to see more different undersampling patterns, which also serves as data augmentation and helps to produce more accurate results. Trainings performed with different orientations mixed together might be able to improve the model's ability to reconstruct different orientations. Furthermore, the ReconResNet3D might perform better handling different orientations than the training orientation - which should be investigated in the future.

The models were only trained with healthy subjects. So, testing the reconstruction performance with volumes containing pathology is an important experiment. In this research, the framework's performance for two different types of pathology was evaluated to judge the clinical applicability. The first type of pathology was Alzheimer disease, which was evaluated using ADNI dataset and yielded close to zero change in BVD (fully-sampled vs reconstructed), which is a good indication and it confirms the network’s ability to preserve the original volume when reconstructing undersampled data. The BVD (fully-sampled vs undersampled) for the Radial 60 spokes was close to the BVD (fully-sampled vs reconstructed) results because unlike Cartesian the radial under-sampling pattern does not distort the image in a way that leads to a change in brain segmentation performed by SIENAX, resulting in the change of brain volume. The second type of pathology that was evaluated here was brain tumours. It was observed qualitatively that the network was able to preserve the tumours while removing most of the artefacts while reconstructing undersampled brain MRIs with tumours from the BraTS dataset. It is to be noted that the resolution of the BraTS dataset is different than the resolution of the training set. Furthermore, the volumes were already brain-extracted while the training set volumes were not. These might have impacted the results. Further experiments should be performed by training the network with healthy brain-extracted volumes and/or experimenting with different brain tumour dataset where brain extraction has not been performed. It is also worth mentioning that an in-depth clinical applicability study, including evaluation by radiologists, was not performed under the scope of this research and will be performed in the future. 

Another interesting thing to point out is that all the trainings and evaluations in this research were conducted using T1 weighted MRIs. However, other MR sequences, such as T2-weighted, FLAIR, etc, are interesting for brain pathologies. The authors hypothesise that the framework should be able to work in a similar fashion with other types of sequences - a hypothesis that will be investigated in the future.

On a closing note, it is to be taken into account that the proposed framework currently works with coil-combined images and the artificial undersampling technique used here only simulates a single-channel data. Real data with multiple channels might produce a better result and can further be accelerated using parallel imaging - this open question is to be investigated in the future.

\section{Conclusion}\label{sec:Conclusion}
A residual learning based framework for undersampled MR data, codenammed NCC1701, has been proposed here. The framework uses the proposed ReconResNet as the network backbone and applies data consistency after reconstructing using the backbone network - making the network model only responsible for predicting the missing data from the undersampled k-space. Evaluation using multiple datasets has shown that the proposed framework can efficiently work with both Cartesian and radial undersampled data, even while trained together, and provides results with high accuracy (SSIM value as high as 0.99) and achieved statistically significant improvements over the baseline methods. Moreover, the proposed framework was able to reconstruct MRIs with pathology (Alzheimer's disease and brain tumours), while being trained using MRIs of only healthy subjects. Experiments presented here have shown that the framework was able to reconstruct properly for undersampled data with an acceleration factor of 20 for Cartesian (2D Varden 5\%) and an acceleration factor of 17 for radial (60 spokes) acquisitions. The acquisition time can further be reduced using techniques such as parallel imaging. Given the fast inference speed of a trained network and the high acceleration factors that can be achieved with the proposed framework, this can be a candidate to be used for real-time MRIs like interventional MRIs. 

\section*{Acknowledgment}

This work was conducted within the context of the International Graduate School MEMoRIAL at Otto von Guericke
University (OVGU) Magdeburg, Germany, kindly supported by the European Structural and Investment Funds (ESF) under the
programme "Sachsen-Anhalt WISSENSCHAFT Internationalisierung" (project no. ZS/2016/08/80646).

\bibliography{mybibfile,related}
\newpage
\section*{Appendix}

\subsection{Dynamics of the learning process: Training vs Validation}
\begin{figure*}[hbt!]
     \centering
     \begin{subfigure}[b]{0.3\textwidth}
         \centering
         \includegraphics[width=\textwidth]{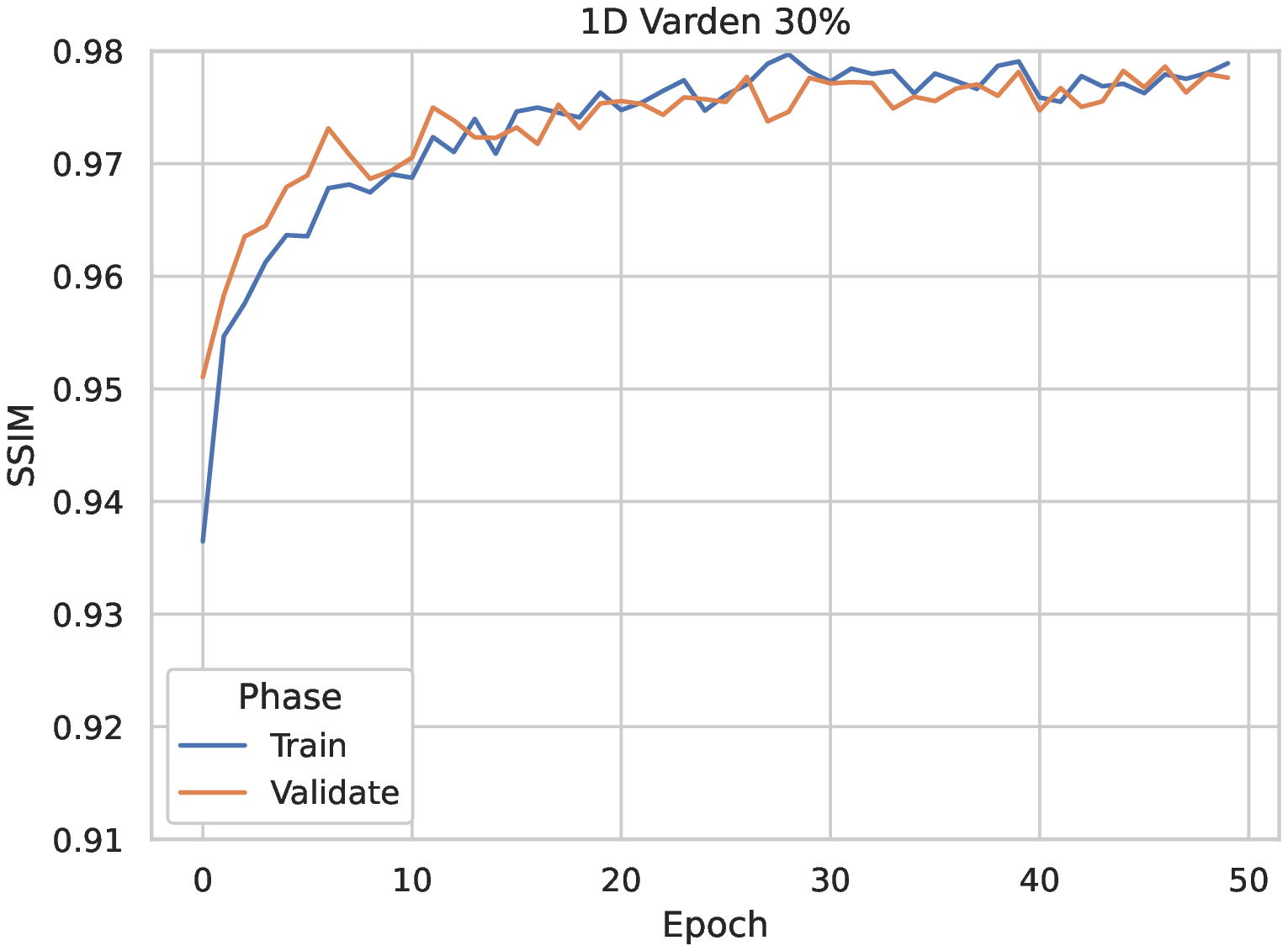}
         \label{fig:trainval_varden1d30}
     \end{subfigure}
     \hfill
     \begin{subfigure}[b]{0.3\textwidth}
         \centering
         \includegraphics[width=\textwidth]{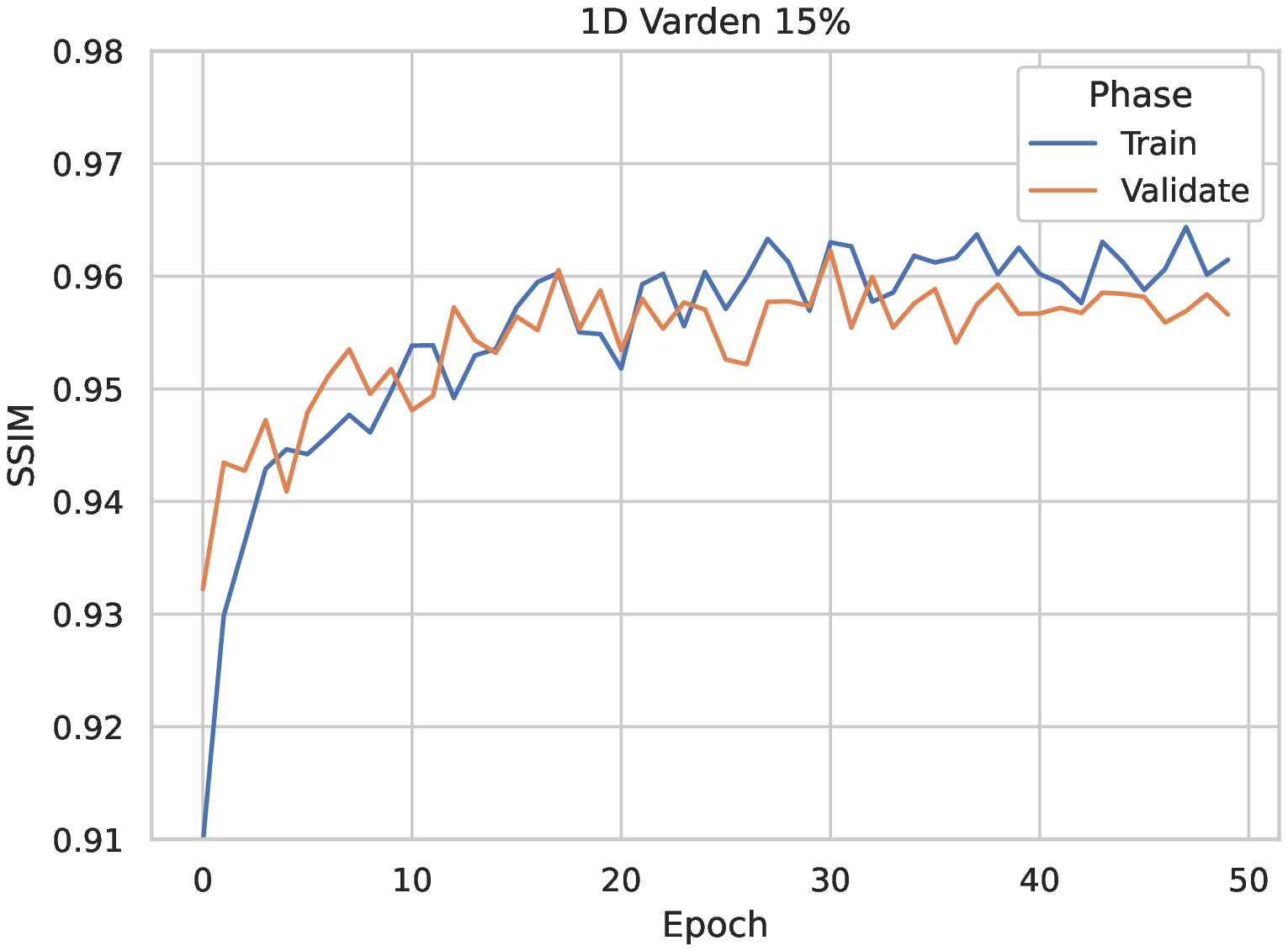}
         \label{fig:trainval_varden1d15}
     \end{subfigure}
     \hfill
     \begin{subfigure}[b]{0.3\textwidth}
         \centering
         \includegraphics[width=\textwidth]{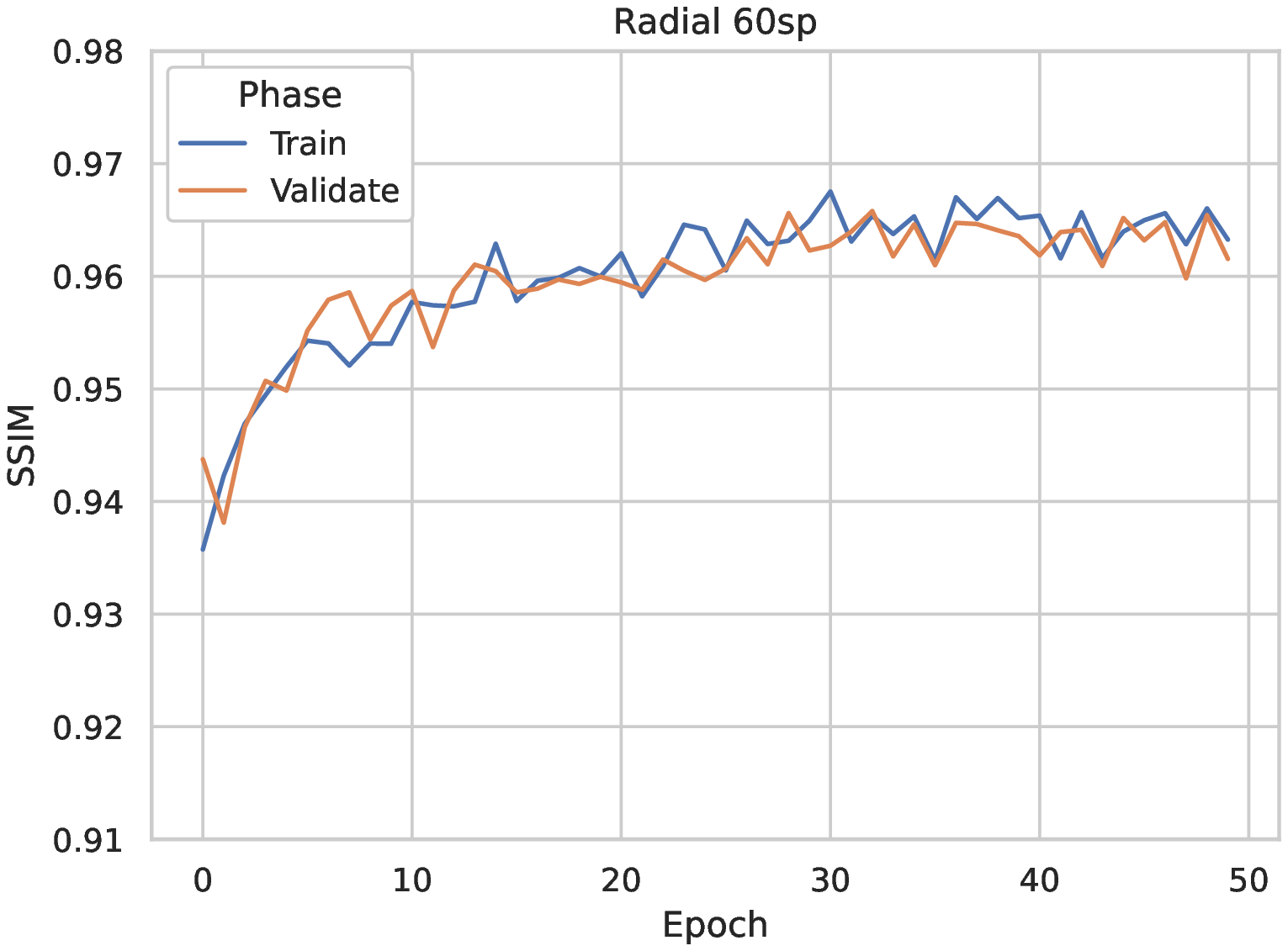}
         \label{fig:trainval_radial60}
     \end{subfigure}
        \caption{Loss curves (SSIM values) during training and validation phase for IXI 3T full volume trainings, for undersamplings 1D Varden 30\% and 15\% of the k-space, radial 60 spokes. Negative of the training loss values were backpropagated during training.}
        \label{fig:trainval}
\end{figure*}

To make sure no overfitting was going on, the training and validation curves were monitored for all the trainings. There was no overfitting been observed during any of the trainings. The curves for IXI 3T full volume trainings are shown in Fig.~\ref{fig:trainval} as examples. Median values were plotted for all the samples present in each epoch. It can be observed that the training and validation curves are indeed very close and show a clean growth through the epochs. Hence, no overfitting behaviour was present.

\subsection{Reconstructing different undersampling patterns than training}

\begin{sidewaysfigure*}
    \centering
    \includegraphics[width=1.1\textwidth]{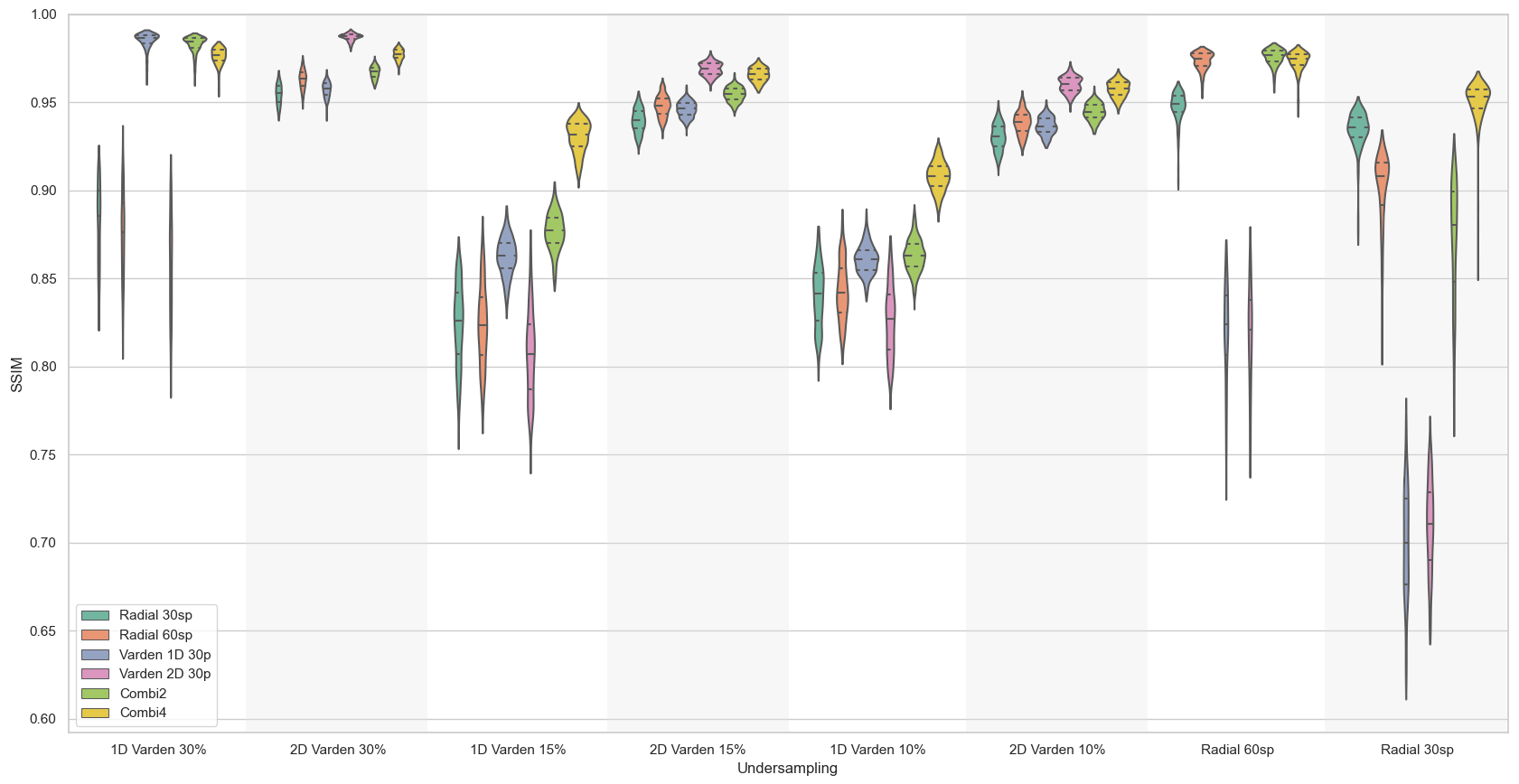}
    \caption{SSIM values for models tested on same or Diﬀerent Pattern, than the training pattern. The x-axis portrays the reconstructed sampling pattern, y-axis portrays the SSIM values, and the different colours portray the different trainings.}
    \label{fig:violin_diffpat}
\end{sidewaysfigure*}

Another tested aspect was the network's ability to handle being tested on different undersampled patterns from the ones used in training. Fig.~\ref{fig:violin_diffpat} presents the resultant SSIM values from these experiments. The x-axis represents individual models and what each is trained on. The y-axis is the SSIM values and the graph's legend consists of the various types of training sets. 

\begin{figure}
\centering
\includegraphics[width=0.48\textwidth]{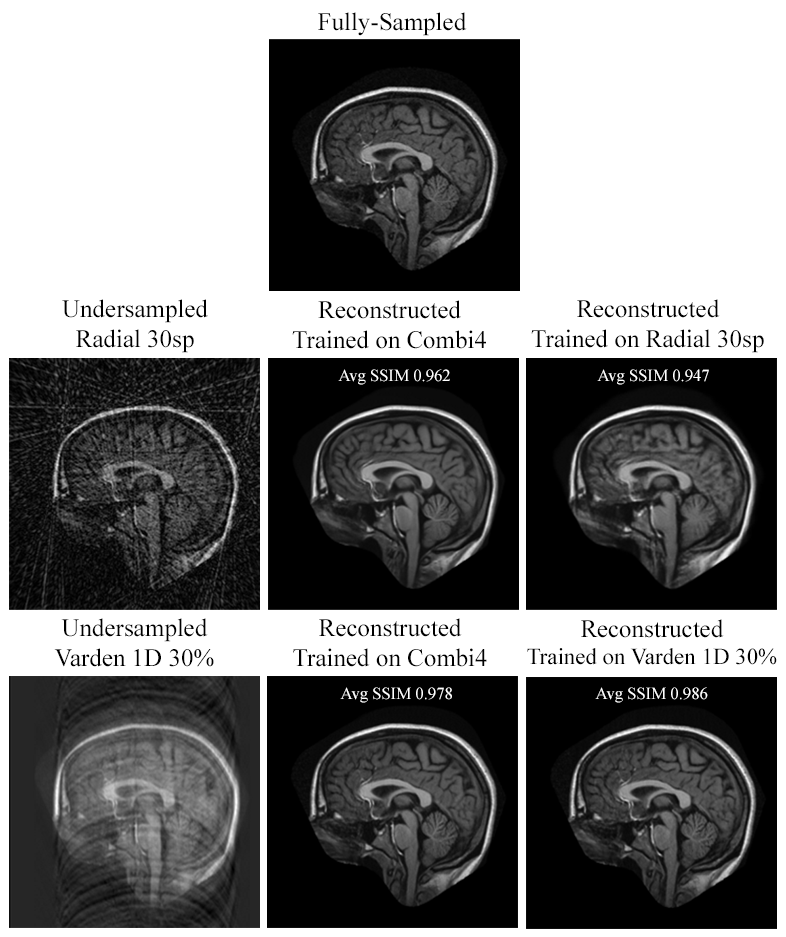}
\caption{Example reconstructions of Radial 30sp and 1D Varden 30\% data: comparing reconstruction quality while being trained on a specific sampling pattern (the one used for reconstruction) or trained on four patterns together (Combi4). Average SSIM values obtained for that specific setup on the whole OASIS dataset are also shown here.}
\label{fig:imgapp_combi4_specific}
\end{figure}

As illustrated in Fig.~\ref{fig:violin_diffpat}, the Combi4 model can be taken as an efficient general model to reconstruct any under-sampling pattern. In the case of reconstructed Radial under-samplings, the Combi4 model was better than the individually trained models of the corresponding under-sampling patterns. In particular, Combi4 had a slightly lower variance than the specific model in reconstructing Radial 60 spokes. Additionally, the mean of the Combi4 model was higher and the variance was better in comparison to the specific model in reconstructing Radial 30 spokes. On the other hand, the plot shows that using a Combi2 model on the same volume yields the same mean but a slightly better variance reconstructing Radial 60 spokes than the specific model. Combi2 models also have the same variance and mean reconstructing Varden 1D 30\% in comparison to its specific model. Fig~\ref{fig:imgapp_combi4_specific} shows an example reconstruction of Radial 30sp and 1D Varden 30\%, where it can be seen that Combi4 is better by 0.015 in reconstructing Radial 30 and, specific Varden 1D 30\% trained model does 0.008 better in estimating the fully-sampled than Combi4 (while comparing the average SSIM values). However, Combi4 model did better on Varden 1D patterns with less than 30\% under-sampling than the Varden 1D 30p trained model (see Fig.~\ref{fig:violin_diffpat}).



\subsection{Examples of 1D Varden 5\% reconstruction}
\begin{figure}[hbt!]
\centering
\includegraphics[width=0.48\textwidth]{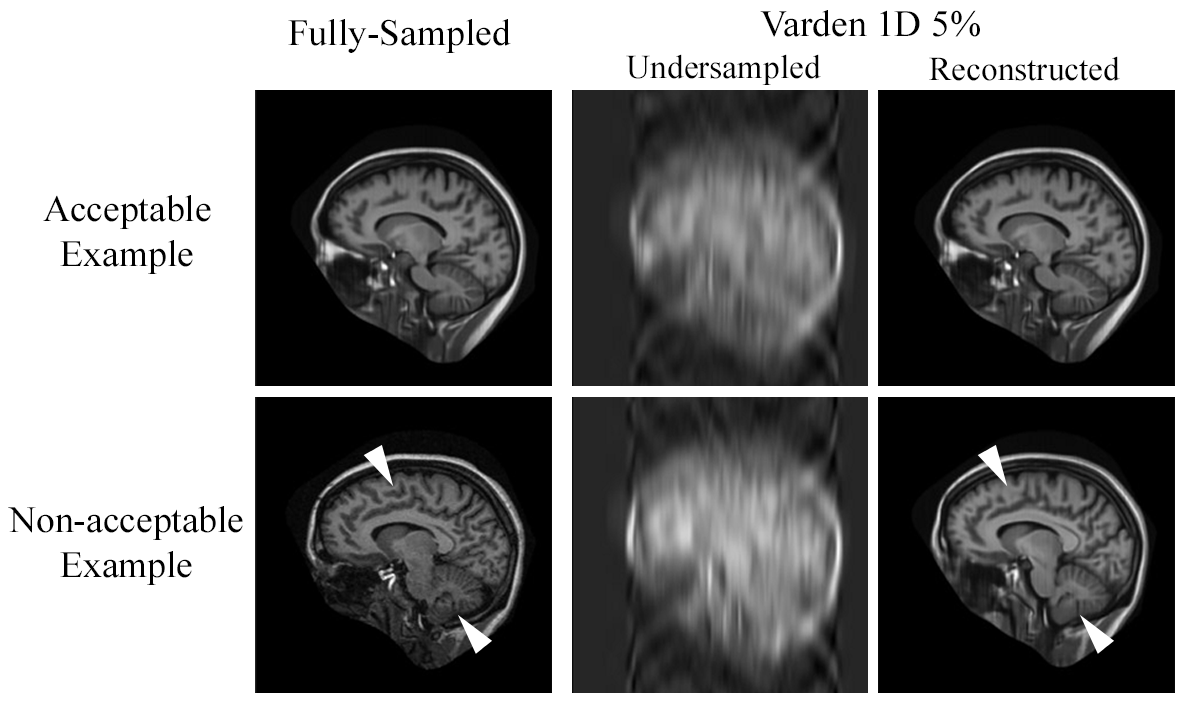}
\caption{The 1D Varden 5\% model outcome, where the fully sampled images on the far left followed by the undersampled and the reconstructed ones. White arrows in the non-acceptable example point to the position where the structural information is lost after reconstruction}
\label{fig:imgapp_varden5p}
\end{figure}

The model 1D Varden 5\% was trained with only a 5\% of the k-space, which may not be sufficient for the network to learn from such sparse information when sampled using 1D Varden sampling. From quantitative analysis (see Sec.~\ref{sec:oasis_limbo}), by observing the distribution of the resultant SSIM values, it can be said that the reconstruction of 1D Varden 5\% might not be reliable. Fig.~\ref{fig:imgapp_varden5p} shows two example reconstructions, both are aesthetically impressive. One of them is highly similar to the ground-truth fully-sampled image. On the other hand in the non-acceptable example, the reconstructed image is smudged and some information is lost. So, it can be said that qualitative analysis is not sufficient and needs to be backed by quantitative analysis.


\end{document}